# Coordinating O&M and Logistical Resources to Enhance Post-Disaster Resilience of Interdependent Power and Natural Gas Distribution Systems

Wei Wang[a], Kaigui Xie[b], Hongbin Wang[c], Tao Chen[c], Hongzhou Chen[c], Yufei He[d]

[a]*School of Electrical Engineering, Southwest Jiaotong University, Pidu, Chengdu 611756, China*
[b]*School of Electrical Engineering, Chongqing University, Shapingba, Chongqing 400044, China*
[c]*Electric Power Research Institute, State Grid Chongqing Electric Power Company, Yubei, Chongqing 401123, China*
[d]*Shinan Power Supply Branch, State Grid Chongqing Electric Power Company, Nan'an, Chongqing 401336, China*

**Abstract**

The increasing interdependence of electric power and natural gas systems, driven by the growth of natural gas-fired generation and the electrification of the gas industry, has underscored the critical need for enhanced resilience to ensure reliable energy supply under adverse conditions, especially in light of recent energy crises. To address this challenge, this paper focuses on interdependent electric power and natural gas distribution systems and proposes a comprehensive strategy to strengthen their resilience, which optimally coordinates diverse resources, encompassing emergency operation and maintenance (O&M) activities and logistical support, to efficiently restore disrupted services and damaged infrastructure. Key measures include dispatching generators, including those with fuel-switching capability, to provide emergency power, supported by the coordinated scheduling of mobile fuel tankers to manage fuel supply logistics. To alleviate operational stresses caused by limited emergency supply, integrated demand response programs for both power and gas are implemented using a zonewise approach, grouping customers with intertwined energy supplies into individual, dispatchable units. Furthermore, the repair of damaged facilities is co-optimized, accounting for variable repair efficiencies influenced by the number and capabilities of dispatched repair units, enabling efficient and coordinated repair activities. The proposed strategy is formulated as a mixed-integer second-order cone programming model and validated through case studies, demonstrating its effectiveness in restoring the interdependent systems and ensuring their continued operation.

**Nomenclature**

***Sets***

$\mathcal{B}$ — Set of branches in EPDS.

$\mathcal{C}(i)/\mathcal{S}(i)$ — Set of EDCs or gas storages supplied by PN $i$.

$\mathcal{F}(i)$ — Set of generators supplied by GN $i$.

$\mathcal{G}_{\mathrm{N/D/dual/ES}}$ — Set of NGFUs/DFUs/dual-fuel units/energy storages.

$\mathcal{G}(i)$ — Set of generators and energy storages at PN $i$.

$\mathcal{M}_{\mathrm{RU}}$ — Set of RUs.

$\mathcal{M}_{\mathrm{GAS/DSL}}$ — Set of gas/diesel tankers.

$\mathcal{N}_{\mathrm{DMG}}$ — Set of damaged facilities.

$\mathcal{N}_{\mathrm{DSL/GAS}}$ — Set of facilities for diesel/gas tankers access.

$\mathcal{N}^{\mathrm{r}}_{\mathrm{DSL}}$ — Set of diesel reservoirs.

$\mathcal{N}^{\mathrm{r}}_{\mathrm{GAS}}/\mathcal{N}^{\mathrm{r}}_{\mathrm{GAS}}(i)$ — Set of gas storages in NGDS, or those located at GN $i$.

$\mathcal{N}_{\mathrm{PN}}/\mathcal{N}_{\mathrm{GN}}$ — Set of power or gas nodes in IENDS.

$\mathcal{N}_{\mathrm{PN/GN.DR}}$ — Set of power or gas nodes related to DR.

$\mathcal{P}$ — Set of pipelines in NGDS. $\mathcal{P} = \mathcal{P}_{\mathrm{pa}} \cup \mathcal{P}_{\mathrm{com}}$, where $\mathcal{P}_{\mathrm{pa}}$ represents passive pipelines and $\mathcal{P}_{\mathrm{com}}$ the EDCs.

$\mathcal{T}$ — Set of time spans for scheduling. $\mathcal{T}=\{1,2,\ldots,D\}$.

$\mathcal{W}(i)$ — Set of gas sources injected into GN $i$.

$\mathcal{Z}_{\mathrm{DR}}$ — Set of zones participating in DR.

***Variables***

$b_{i,t,y}$ — Binary variable. 1 if $y$ RUs are assigned to repair facility $i$ during TS $t$, 0 otherwise.

$C_{\mathrm{N/D},k,t}$ — Natural gas or diesel consumption of generator $k$ during TS $t$.

$D_{j,i,t}/N_{j,i,t}$ — Fuel output from diesel or gas tanker $j$ to storage $i$ during TS $t$.

$F_{\mathrm{DR},(i,j),t}$/ $F_{\mathrm{DR},i,t}$ Gas demand reduction at zone $(i, j)$ or GN $i$ resulting from DR during TS $t$.

$F_{ii',t}$ Gas flow through pipeline $(i, i')$ from GN $i$ to $i'$ during TS $t$.

$L_{k,t}$ Gas flow that NGFU or dual-fuel unit $k$ absorbs from NGDS during TS $t$.

$L_{\mathrm{stg},k,t}$ Gas flow that gas storage $k$ releases to NGDS.

$L_{\mathrm{src},k,t}$ Output flow from gas source $k$ during TS $t$.

$L_{\mathrm{inj/wd},i,t}$ Gas flow injected into or withdrawn from gas storage $i$ by the NGDS during TS $t$.

$P_{k,t}$/$Q_{k,t}$ Active or reactive power output of generator or energy storage $k$ during TS $t$.

$P_{\mathrm{N/D},k,t}$ Active power output of dual-fuel unit $k$ in the gas or diesel mode during TS $t$.

$P_{ii',t}$/$Q_{ii',t}$ Active or reactive power flow in branch $(i, i')$ from PN $i$ to $i'$ during TS $t$.

$P_{\mathrm{ch},k,t}$/$P_{\mathrm{dch},k,t}$ Charging or discharging power of energy storage $k$ during TS $t$.

$P_{\mathrm{com/stg},k,t}$ Power demand of EDC or gas storage $k$ during TS $t$.

$P_{\mathrm{DR},(i,j),t}$/ $Q_{\mathrm{DR},(i,j),t}$, $P_{\mathrm{DR},j,t}$/$Q_{\mathrm{DR},j,t}$ Active or reactive power demand reduction at zone $(i, j)$ and at PN $j$ resulting from DR during TS $t$, respectively.

$soc_{k,t}$ State of charge of energy storage $k$ at the end of TS $t$.

$u_{i,t}^2$ Squared voltage magnitude at PN $i$ during TS $t$.

$x_{j,i,t}$/ $v_{j,i,t}$ Binary variable. 1 if mobile resource $j$ (whether a RU or fuel tanker) is parked at/traveling to facility $i$ during TS $t$, 0 otherwise.

$y_{ii',t}$, $Y_{ii',t}$ Auxiliary binary and continuous variables to convexify Weymouth equations.

$z_{\mathrm{N/D},k,t}$ Binary variable. 1 if dual-fuel unit $k$ runs on natural gas/diesel during TS $t$, 0 otherwise.

$\alpha_{\mathrm{ch/dch,k,t}}$ Binary variable. 1 if energy storage $k$ charges/discharges during TS $t$.

$\alpha_{\mathrm{inj/wd},i,t}$ Binary variable. 1 if NGDS injects into or withdraws from gas storage $i$ during TS $t$.

$\gamma_{\mathrm{P/N},(i,j),t}$ Binary variable. 1 if power/gas DR is executed at zone $(i, j)$, 0 otherwise.

$\delta_{\mathrm{P/N},i,t}$ Binary variable. 1 if power/gas load at PN/GN $j$ is picked up during TS $t$, 0 otherwise.

$\tau_{\mathrm{N/D},l,k,t}$ Binary variable. 1 if generator $k$ operates within the $l$th segment of power-fuel consumption curve when burning natural gas/diesel during TS $t$, 0 otherwise.

$\eta_{i,t}$ Repair efficiency for facility $i$ during TS $t$.

$\kappa_{i,t}$ Binary variable. 1 if facility $i$ is operational during TS $t$, 0 otherwise.

$\lambda_{k,t}$ Binary variable. 1 if dual-fuel unit $k$ switches its fuel from TS $t-1$ to $t$.

$\Lambda_{k,t}$ Binary variable. 1 if compressor $k$ works, 0 otherwise.

$\pi_{i,t}^2$ Squared pressure at GN $i$ during TS $t$.

$\chi_{j,i,t}$ Binary variable. 1 if fuel output from fuel tanker $j$ to facility $i$ is allowed during TS $t$.

$\psi_{i,t}$ The amount of fuel in facility $i$ (whether a fuel tanker or a fixed-location storage) at the end of TS $t$.

***Parameters***

$a_{\mathrm{N/D},l,k}$, $b_{\mathrm{N/D},l,k}$ Coefficients to linearize the power-fuel consumption curve of generator $k$ when burning natural gas/diesel.

$D_{j,\mathrm{in,max}}$/ $D_{j,\mathrm{out,max}}$ Maximum fuel input or output of diesel tanker $j$ within a TS.

$e_{\mathrm{inj/wtd},i}$ Efficiency of injecting gas to or withdrawing gas from gas storage $i$.

$e_{\mathrm{ch/dch},k}$ Charging/discharging efficiency of energy storage $k$.

$E_k$ Energy capacity of energy storage $k$.

$F_{ii',\mathrm{max}}$ Transmission capacity of pipeline $(i, i')$.

$H_{(i,j)}$ Upper limit for integrated power and gas DR.

$K_{ii'}$ Weymouth constant of pipeline $(i, i')$.

$L_{\mathrm{N/D},k}$ The number of segments in the modeling of power-fuel consumption curve of generator $k$ when burning natural gas or diesel.

$L_{\mathrm{inj/wd},i,\mathrm{max}}$ Maximum gas flow that NGDS can inject into or withdraw from gas storage $i$.

$L_{\mathrm{src},k,\mathrm{min/max}}$ Lower or upper limit of output from gas source $k$.

$N_{j,\mathrm{in,max}}$/ $N_{j,\mathrm{out,max}}$ Maximum fuel input to or output from gas tanker $j$ within a TS.

$p_{\mathrm{N/D},l,k}$ Upper limit of active power output within the $l$th segment of power-fuel consumption curve of generator $k$ when burning natural gas or diesel.

$p_{\mathrm{N/D,max},k}$ Upper limit of active power output of dual-fuel unit $k$ when burning natural gas or diesel.

$P_{\mathrm{ch/dch},k,\mathrm{max}}$ Maximum charging or discharging power of energy storage $k$.

$\tilde{P}_{(i,j),t}$/$\tilde{F}_{(i,j),t}$ Original active power or gas demand of zone $(i, j)$ w/o DR.

$\tilde{P}_{i,t}(\tilde{Q}_{i,t})$/$\tilde{F}_{i,t}$ Original active (reactive) power or gas demand of PN/GN $i$ w/o DR.

$Q_{k,\mathrm{max}}$ Maximum reactive power output from generator or energy storage $k$.

$r_{ii'}$/$x_{ii'}$ Resistance or reactance of branch $(i, i')$.

$S_{k,\mathrm{app}}$ Apparent capacity of generator or energy storage $k$.

| | |
|---|---|
| $S_{(i,i'),\mathrm{app}}$ | Apparent capacity of branch ($i$, $i'$). |
| $soc_{k,\min}$/ $soc_{k,\max}$ | Lower or upper limit of state of charge of energy storage $k$. |
| $TP_{(i,j),\max}$/ $TN_{(i,j),\max}$ | Maximum allowable duration of power or gas DR events in zone ($i$, $j$) over the entire scheduling horizon. |
| $TP_{(i,j),\mathrm{du},\max/\min}$/$TN_{(i,j),\mathrm{du},\max/\min}$ | Maximum/minimum allowable duration of a single power or gas DR event in zone ($i$, $j$). |
| $TP_{(i,j),\mathrm{int},\min}$/ $TN_{(i,j),\mathrm{int},\min}$ | Minimum allowable interval between two consecutive power or gas DR events in zone ($i$, $j$). |
| $u^2_{i,\min}$/$u^2_{i,\max}$ | Lower or upper limit of the squared voltage magnitude at PN $i$. |
| $w_{\mathrm{P/N},i}$ | Weight of load at PN/GN $i$. |
| $\beta_{i,y}$ | Repair efficiency with $y$ RUs repairing facility $i$. |
| $\zeta_{k,\max}$/$\zeta_{k,\min}$ | Maximum or minimum compression ratio of compressor $k$. |
| $\pi^2_{i,\min}$/$\pi^2_{i,\max}$ | Lower or upper limit of the squared pressure at GN $i$. |
| $\rho_{\mathrm{com},k}$ | Coefficient for calculating the power consumption of compressor $k$. |
| $\rho_{\mathrm{stg,inj},i}$/ $\rho_{\mathrm{stg,wd},i}$ | Coefficient for calculating the power consumption of gas storage $i$. |
| $\varsigma_{1/2}$, $o_{1/2/3/4}$ | Coefficients in the objective function. |
| $\sigma_{\mathrm{P/N},(i,j),\min/\max}$ | Minimum or maximum ratio of demand reduction resulting from power or gas DR in zone ($i$, $j$). |
| $\varphi_{(i,j)}$/$\varphi_i$ | Power factor of zone ($i$, $j$) or facility $i$. |
| $\psi_{i,max}$ | Fuel storage capacity of facility $i$. |

## 1. Introduction

Recent decades have witnessed an unprecedented rise in the interdependence between electricity and natural gas infrastructures, a trend likely to persist [1]. Natural gas-fired generating units (NGFUs) have become increasingly favored in the power sectors due to low costs and carbon emissions, making natural gas a major electricity source. The IEA reports that in 2023, natural gas provided about a quarter of global electricity generation, with 40% of gas consumed for this purpose [2]. On the other hand, the gas sectors depend on electricity for essential operations in gas production, processing, transmission, and distribution. Driven by global emission goals, electrification is gaining momentum in the gas industry [3], exemplified by the growing preference for electricity-driven compressors (EDCs), which offer significant environmental benefits over traditional ones [4], [5]. The Southwest Energy Efficiency Project notes that replacing 200 large natural gas engines in compressors with electric motors by 2030 could reduce GHG emissions by 45.2% [6]. This shift not only reduces the environmental footprint of the gas industry but also increases its electricity dependence.

While the interdependence between power and gas systems supports daily operations, it poses more critical challenges in emergencies, with closely linked impacts and responses to disasters affecting both systems. This can be exemplified by the aggravated risk of cascading and systemic failures [7], [8]. For example, a decline or interruption in the natural gas supply to NGFUs, resulting from freeze-related shutdowns of gas wellheads or increased demand for gas heating, can reduce or halt their power generation. This, in turn, may lead to a loss of power supply to electricity-powered gas facilities like EDCs, further impacting gas flows, as seen in the 2011 and 2021 cold weather events in the USA [5], [9]. Just as in the cascading failure process, the restoration processes are also reciprocal, with the recovery of one system's service often facilitating another's recovery through their interdependence or coupling components [10]. While preventing disasters may be challenging and costly, efforts can be directed towards efficiently restoring the impacted power and gas systems, ensuring systems are more resilient and homes remain lit and warm in the face of future similar events.

Research has increasingly focused on enhancing the resilience of power and gas systems following disasters, particularly targeting the distribution systems that are most susceptible to failures. Among the measures explored, distributed generation (DG) has proven especially effective in restoring service of electric power distribution systems (EPDS) when they or a part of them lose connection to the main grid, by intentionally forming islands around DG [11], [12], [13]. However, while significant focus has been placed on dispatching DG for this purpose, the critical issue of fuel supply for fuel-based DGs is often overlooked. Although the pre-allocation of fuel resources for EPDS resilience enhancement has been addressed in some studies [14], the sudden and unforeseeable nature of many disasters presents a major challenge in consistently ensuring sufficient pre-deployed fuel for backup DG units, thereby jeopardizing their operational readiness during emergencies. Furthermore, given the interdependence between power and gas systems, while DGs like NGFUs can access gas from robust underground pipelines, this supply is not always reliable and may be restricted due to operational stresses within the gas system. For example, as seen in past energy crises, the supply to NGFUs may be reduced or curtailed to prioritize essential customers, such as residences. During the 2021 event [9], 31.4% of all unplanned generating unit outages were caused by fuel issues, with 87% of those outages attributed to natural gas supply issues. Therefore, ensuring adaptive and well-coordinated fuel allocation alongside generator dispatching is crucial for effective restoration efforts. However, this critical issue remains insufficiently addressed in existing research, highlighting a notable gap in restoration strategies involving the utilization of DGs.

Compared to EPDS, natural gas distribution systems (NGDS) generally exhibit greater resilience to many extreme disasters due to their underground layout [15], except in cases of extreme cold weather, which can lead to pipeline freeze-offs. In the event of gas shortages, which may stem from upstream supply shortfalls caused by freeze-offs, outages of EDCs, or increased demand, gas storages can play a critical role in maintaining an adequate supply to customers [16], [17]. In this scenario, they function similarly to DGs in providing backup supply during emergencies, yet both operations are conditional: just as DG requires fuel to run, gas storage may also rely on electricity to facilitate gas exchanges with NGDS, akin to the EDC.

As previously stated, the interdependent electric power and natural gas distribution systems (IENDS) exhibit cascading impacts and reciprocal restoration behaviors in response to disasters. Analyzing these systems in isolation fails to capture their intricate interdependencies, highlighting the need to treat them as a unified system for effective reliability- or resilience-focused modeling, analysis, and decision-making [18], [19]. In light of this, and driven by the growing focus on resilience, research has been emerging on the cooperative restoration of IENDS through the strategic use of measures such as those aforementioned. For example, in [20], the operation of DGs and their energizing paths to loads were optimized to restore the IENDS. In [21], the supplying paths to gas loads through controllable valves were further co-optimized. Generators and gas line packs were utilized in [22] to support the IENDS with emergency power and gas supplies. Additionally, [23] developed a comprehensive strategy for service restoration of the IENDS after earthquakes, involving more resources such as generators, electricity and gas storages, and hydrogen production. However, while existing works emphasize the strategic utilization of supply-side resources to facilitate the restoration of IENDS, there remains rather insufficient exploration into the potential and integration of demand-side measures, as well as their further coordination with repair efforts.

While essential emergency supplies can be provided by DG and gas storage, they are typically temporary and constrained due to factors such as DG unit capacity, available fuel, and storage volume. In such emergencies, it is critical for both IENDS operators and customers to share the responsibility and collaborate to ensure the continuity and swift restoration of energy services. To alleviate such operational stresses and help balance the limited supply, demand response (DR) can be implemented by temporarily reducing demands, either for power or gas, until the affected systems overcome the stressed condition. In this regard, power DR has proven effective in supporting the restoration of the EPDS in a few studies, whether it is implemented at normal feeders to free up extra capacity for picking up more interrupted loads [24], or collaborates with DG to either accelerate the restoration or enhance the subsequent emergency energy management for isolated EPDS, thereby extending its survival time [25], [26].

DR has also been explored in the gas industry in recent years. Interruptible gas customers can be curtailed during peak hours or emergencies, a practice used in several events [5], [9], [16]. In addition to this traditional approach, some pilot DR programs are being launched to address growing seasonal stress on gas systems [27]. Programs by National Grid, ConEd, and SoCalGas, for instance, allow utilities to remotely switch customers' gas loads to alternative fuels or adjust their thermostat settings [28], known as direct load control (DLC), which has long been used in the electric industry. According to [29], a 12°F overnight setback on a home's thermostat can bring up to an 8% reduction in gas consumption, and the reduction could reach almost 30% with the setpoint consistently set back to 63°F. However, compared to the well-developed power DR, gas DR has received much less attention and remains in its early stages of research. Although a limited number of studies have emerged on the integrated dispatching of power and gas DRs, almost all of them focus on normal conditions, such as aiming towards a low-carbon and economic operation of the interdependent power and gas systems [30], [31]. During emergencies, the urgent need to balance limited supply and demand highlights the significant, yet untapped, potential of power and gas DRs, especially their integration, to support the restoration and emergency operations of the interdependent systems. In addition, existing studies primarily focus on dispatching DR at the node level, or even down to individual customers. While both approaches are useful, they face challenges due to the intertwined supply relationships among customers, which will be discussed in detail later.

In addition, repair activities are among the most crucial and direct means during the restoration phase, and a few studies have explored the coordinated scheduling of repair activities with other emergency resources. For example, ref. [32] provides a model for the repair process based on the combination of continuous- and discrete-time formulations, thereby improving the IENDS resilience through coordination with resources like DGs. Ref. [33] optimizes repair activities for the traction power supply system, considering the effects of network reconfiguration and cross-border power supply as emergency resources. Ref. [34] proposes a comprehensive method for optimizing the post-disaster scheduling of repair activities alongside various types of DGs to enhance EPDS resilience. However, current models of repair activities have yet to fully harness their potential, often relying on a pre-defined constant repair time. In practice, this time may vary due to acceleration arising from cooperation among repair crews, making it a variable dependent on the dispatched crews and allowing for greater flexibility in repair decision-making.

In light of the research gaps, we propose a comprehensive strategy to enhance the post-disaster resilience of the IENDS. This strategy coordinates various resources to restore impacted services, including O&M resources that cover supply-side, demand-side, and repair efforts, as well as the logistical fuel supply to DGs. The main contributions are summarized as follows:

*1)* Fuel supply issues are particularly addressed for various DG units through the dispatching of mobile fuel tankers. The fuel delivery processes are decomposed and modeled to facilitate the coordinated scheduling of fuel deployment and generator operations.

*2)* Integrated power and gas DRs are incorporated into the restoration phase of the IENDS and dispatched to alleviate stressed conditions, with a zonewise approach strategically grouping customers into distinct, dispatchable DR units, thereby ensuring the practicality and accuracy of DR scheduling.

*3)* Additionally, the dispatching of mobile repair units (RUs) is modeled to enable coordinated repair activities during restoration. Variable repair efficiencies are incorporated, offering flexibility in the decision-making for repair: this allows RUs to work independently on parallel repairs or collaborate for accelerated repairs, thereby enhancing the overall repair efficiency.

## 2. Post-disaster Restoration Process of IENDS

Fig. 1 illustrates the proposed strategy. After disasters, the IENDS may sustain extensive damages, particularly targeting the exposed EPDS component. To restore the degraded services, DGs, including NGFUs, diesel-fired units (DFUs), dual-fuel units, and energy storages, will offer backup power supplies, while natural gas storages, assumed to rely on electricity to operate, can support gas supplies. Fuel for DG generation will be guaranteed by dispatching mobile fuel tankers between the units and fuel reservoirs. Meanwhile, power and gas DRs will be properly implemented for dispersed customers to address operational stresses, for example, those resulting from limited supplies. Additionally, the mobile RUs will be dynamically dispatched to repair damaged facilities, offering a permanent solution to restore the access for normal and stable supplies.

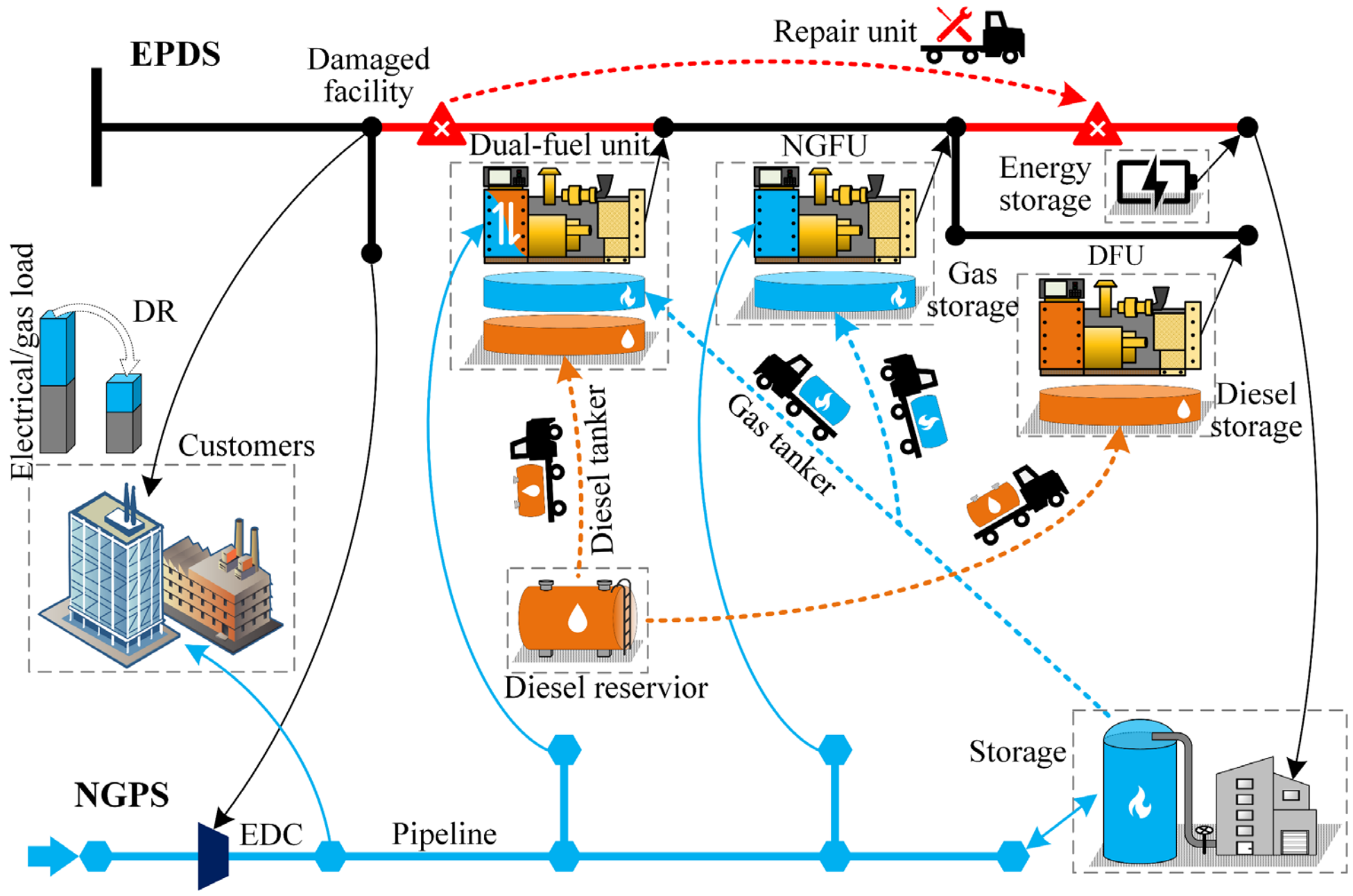

Fig. 1. How the strategy works to facilitate IENDS restoration.

## 3. Mathematical Formulations

In this section, we construct the model for the proposed strategy. Specifically, the mobility model proposed in [35] is used for the routing of mobile resources, including fuel tankers and RUs. For brevity, this model is not provided here.

### *3.1. Modeling of Fuel Delivery for Generators*

#### *3.1.1. Fuel consumption of NGFU/DFU*

Manufacturers often provide fuel consumption data for their generators at typical load levels. With these breakpoints, we use piecewise linearization to model the relationship between power output and fuel consumption. Specifically, for NGFUs:

$$\begin{gathered} p_{\mathrm{N},l-1,k} - P_{k,t} \le M\left(1-\tau_{\mathrm{N},l,k,t}\right) \;\;,\;\; P_{k,t} - p_{\mathrm{N},l,k} \le M\left(1-\tau_{\mathrm{N},l,k,t}\right) \;\;, \\ \forall k \in \mathcal{G}_{\mathrm{N}}, t \in \mathcal{T}, l \in \left\{1, 2, \cdots, L_{\mathrm{N},k}\right\} \end{gathered} \tag{1a}$$

$$\begin{gathered} C_{\mathrm{N},k,t} - \left(a_{\mathrm{N},l,k} P_{k,t} + b_{\mathrm{N},l,k}\right) \le M\left(1-\tau_{\mathrm{N},l,k,t}\right), \\ -M\left(1-\tau_{\mathrm{N},l,k,t}\right) \le C_{\mathrm{N},k,t} - \left(a_{\mathrm{N},l,k} P_{k,t} + b_{\mathrm{N},l,k}\right), \\ \forall k \in \mathcal{G}_{\mathrm{N}}, t \in \mathcal{T}, l \in \left\{1, 2, \cdots, L_{\mathrm{N},k}\right\} \end{gathered} \tag{1b}$$

$$\sum_{l=1}^{L_k} \tau_{\mathrm{N},l,k,t} = 1 \;\;,\;\; \forall k \in \mathcal{G}_{\mathrm{N}}, t \in \mathcal{T} \tag{1c}$$

Additionally, similar constraints can be created for DFUs.

#### *3.1.2. Fuel consumption of dual-fuel unit*

Many dual-fuel units operate on gas but can switch to diesel when needed, providing extra fuel adequacy, as highlighted by past winter events [9]. By 2021, units with such fuel-switching capability accounted for about 18% of U.S. generating capacity, ranging from several hundred kW to MW [36]. To model their fuel consumption under both fuel modes, we have:

$$z_{\mathrm{N},k,t} + z_{\mathrm{D},k,t} = 1 \;\;,\;\; \forall k \in \mathcal{G}_{\mathrm{dual}}, t \in \mathcal{T} \tag{2a}$$

$$
\begin{aligned}
& 0 \le P_{\mathrm{N},k,t} \le z_{\mathrm{N},k,t} p_{\mathrm{N},\max,k} \quad , \quad 0 \le C_{\mathrm{N},k,t} \le z_{\mathrm{N},k,t} M \quad , \\
& 0 \le P_{\mathrm{D},k,t} \le z_{\mathrm{D},k,t} p_{\mathrm{D},\max,k} \quad , \quad 0 \le C_{\mathrm{D},k,t} \le z_{\mathrm{D},k,t} M \quad , \\
& \forall k \in \mathcal{G}_{\mathrm{dual}} , t \in \mathcal{T}
\end{aligned} \tag{2b}
$$

$$
\begin{aligned}
& \lambda_{k,t} \ge z_{\mathrm{N},k,t-1} - z_{\mathrm{N},k,t} \quad , \quad \lambda_{k,t} \ge z_{\mathrm{D},k,t-1} - z_{\mathrm{D},k,t} \quad , \\
& \forall k \in \mathcal{G}_{\mathrm{dual}} , t \in \mathcal{T} \setminus \{1\}
\end{aligned} \tag{2c}
$$

$$
\sum_{t \in \mathcal{T} \setminus \{1\}} \lambda_{k,t} \le \lambda_{k,\max} \quad , \quad \forall k \in \mathcal{G}_{\mathrm{dual}} \tag{2d}
$$

Constraint (2a) defines the two fuel modes, and (2b) bounds the power output and fuel consumption for each. Constraint (2c) and (2d) permit the switching a limited number of times [37]. Similar constraints to (1) should be created for both fuel modes, but due to space limits, they are not detailed here.

### *3.1.3. Fuel exchanges among facilities*

Assuming all units have onsite fuel storage to receive gas or diesel from fuel tankers, and units at the same location share the same storage. Fuel exchanges among facilities, including fuel tankers, onsite storage beside the units, and fuel depots such as diesel reservoirs and gas storages in the NGDS, are illustrated in Fig.2 and modeled by (3).

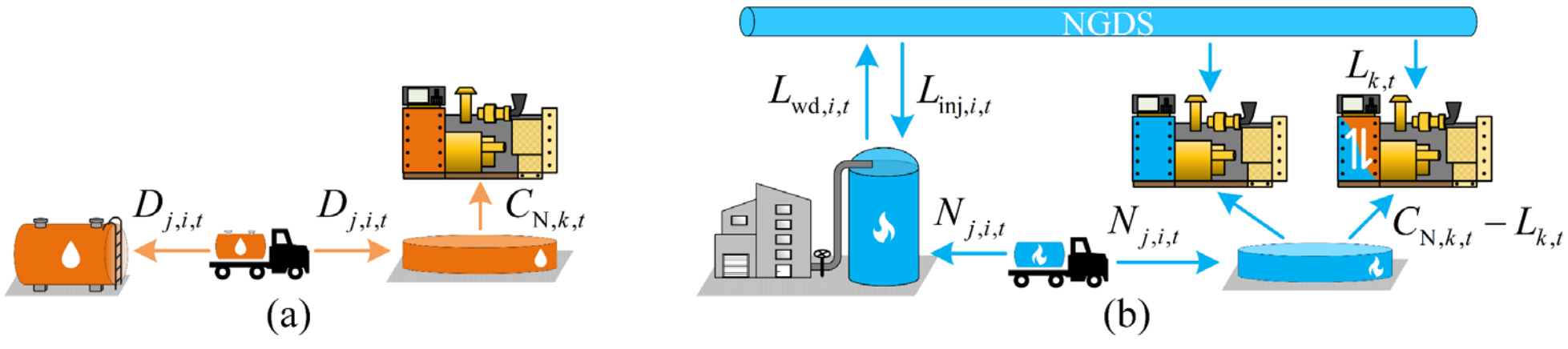

Fig. 2. Fuel exchanges among (a) diesel and (b) gas facilities.

Specifically, for diesel tankers, we have:

$$
\psi_{j,t} = \psi_{j,t-1} - \sum_{i \in \mathcal{N}_{\mathrm{DSL}}} D_{j,i,t} \quad , \quad \forall j \in \mathcal{M}_{\mathrm{DSL}} , t \in \mathcal{T} \tag{3a}
$$

$$
\begin{aligned}
& \chi_{j,i,t} \le x_{j,i,t} \quad , \quad -D_{j,\mathrm{in},\max} \chi_{j,i,t} \le D_{j,i,t} \le D_{j,\mathrm{out},\max} \chi_{j,i,t} \quad , \\
& \forall j \in \mathcal{M}_{\mathrm{DSL}} , i \in \mathcal{N}_{\mathrm{DSL}} , t \in \mathcal{T}
\end{aligned} \tag{3b}
$$

Fuel exchange for diesel reservoirs can be formulated by

$$
\psi_{i,t} = \psi_{i,t-1} + \sum_{j \in \mathcal{M}_{\mathrm{DSL}}} D_{j,i,t} \quad , \quad \forall i \in \mathcal{N}_{\mathrm{DSL}}^{\mathrm{r}} , t \in \mathcal{T} \tag{3c}
$$

Then, for onsite diesel storage beside DFU or dual-fuel unit, fuel exchange with tankers and units is formulated by:

$$
\begin{aligned}
& \psi_{i,t} = \psi_{i,t-1} + \sum_{j \in \mathcal{M}_{\mathrm{DSL}}} D_{j,i,t} - \sum_{k \in \mathcal{G}_{\mathrm{D}}(i) \cup \mathcal{G}_{\mathrm{dual}}(i)} C_{\mathrm{D},k,t} \quad , \\
& \forall i \in \mathcal{N}_{\mathrm{DSL}} \setminus \mathcal{N}_{\mathrm{DSL}}^{\mathrm{r}} , t \in \mathcal{T}
\end{aligned} \tag{3d}
$$

Similarly, for gas tankers, we have:

$$
\psi_{j,t} = \psi_{j,t-1} - \sum_{i \in \mathcal{N}_{\mathrm{GAS}}} N_{j,i,t} \quad , \quad \forall j \in \mathcal{M}_{\mathrm{GAS}} , t \in \mathcal{T} \tag{3e}
$$

$$
\begin{aligned}
& \chi_{j,i,t} \le x_{j,i,t} \quad , \quad -N_{j,\mathrm{in},\max} \chi_{j,i,t} \le N_{j,i,t} \le N_{j,\mathrm{out},\max} \chi_{j,i,t} \quad , \\
& \forall j \in \mathcal{M}_{\mathrm{GAS}} , i \in \mathcal{N}_{\mathrm{GAS}} , t \in \mathcal{T}
\end{aligned} \tag{3f}
$$

For gas storages in the NGDS, they can inject and withdraw gas from the NGDS and refuel gas tankers, as formulated by (3g) and (3h). Compared to tankers or onsite storages, the much larger gas storages in the NGDS generally exhibit lower injection and withdrawal efficiency due to their more complex infrastructure, leading to increased gas leakage. Therefore, this efficiency is considered in the modeling.

$$\psi_{i,t}=\psi_{i,t-1}+\sum_{j\in\mathcal{M}_{\mathrm{GAS}}}N_{j,i,t}+\left(e_{\mathrm{inj},i}L_{\mathrm{inj},i,t}-L_{\mathrm{wd},i,t}/e_{\mathrm{wd},i}\right)\Delta t\quad, \\ \forall i\in\mathcal{N}_{\mathrm{GAS}}^{\mathrm{r}},t\in\mathcal{T} \tag{3g}$$

$$\alpha_{\mathrm{inj},i,t}+\alpha_{\mathrm{wd},i,t}\le 1\quad,\quad 0\le L_{\mathrm{inj},i,t}\le\alpha_{\mathrm{inj},i,t}L_{\mathrm{inj},i,\max}\quad, \\ 0\le L_{\mathrm{wd},i,t}\le\alpha_{\mathrm{wd},i,t}L_{\mathrm{wd},i,\max}\quad,\quad L_{\mathrm{stg},i,t}=L_{\mathrm{wd},i,t}-L_{\mathrm{inj},i,t}\quad, \\ \forall i\in\mathcal{N}_{\mathrm{GAS}}^{\mathrm{r}},t\in\mathcal{T} \tag{3h}$$

Fuel exchange for onsite gas storage beside NGFUs or dual-fuel units is similar to (3d). Additionally, these units typically have pipelines for direct supply from NGDS. Thus, we have:

$$\psi_{i,t}=\psi_{i,t-1}+\sum_{j\in\mathcal{M}_{\mathrm{GAS}}}N_{j,i,t}-\sum_{k\in\mathcal{G}_{\mathrm{N}}(i)\cup\mathcal{G}_{\mathrm{dual}}(i)}\left(C_{\mathrm{N},k,t}-L_{k,t}\Delta t\right), \\ \forall i\in\mathcal{N}_{\mathrm{GAS}}\setminus\mathcal{N}_{\mathrm{GAS}}^{\mathrm{r}},t\in\mathcal{T} \tag{3i}$$

$$0\le L_{k,t}\Delta t\le C_{\mathrm{N},k,t}\quad,\quad\forall k\in\mathcal{G}_{\mathrm{N}}\cup\mathcal{G}_{\mathrm{dual}},t\in\mathcal{T} \tag{3j}$$

Finally, fuel storage capacities of facilities are limited by:

$$0\le\psi_{i,t}\le\psi_{i,\max}\;,\forall i\in\mathcal{M}_{\mathrm{DSL}}\cup\mathcal{N}_{\mathrm{DSL}}\cup\mathcal{M}_{\mathrm{GAS}}\cup\mathcal{N}_{\mathrm{GAS}},t\in\mathcal{T} \tag{3k}$$

### *3.2. Modeling of Operation for DGs*

The power outputs of NGFUs are constrained by (4a) and (4b). Note that the active power is already limited by (1).

$$0\le Q_{k,t}\le Q_{k,\max}\quad,\quad\forall k\in\mathcal{G}_{\mathrm{N}},t\in\mathcal{T} \tag{4a}$$

$$P_{k,t}^2+Q_{k,t}^2\le S_{k,\mathrm{app}}^2\quad,\quad\forall k\in\mathcal{G}_{\mathrm{N}},t\in\mathcal{T} \tag{4b}$$

The operation of DFUs and dual-fuel units (for each mode) can be similarly formulated. Constraint (4b) and similar ones in the following sections can be linearized using the method described in [38]. Additionally, for dual-fuel units, (4c) is added to express their actual active power output.

$$P_{k,t}=P_{\mathrm{N},k,t}+P_{\mathrm{D},k,t}\quad,\quad\forall k\in\mathcal{G}_{\mathrm{dual}},t\in\mathcal{T} \tag{4c}$$

Furthermore, for energy storages we have:

$$soc_{k,t}=soc_{k,t-1}+\left(e_{\mathrm{ch},k}P_{\mathrm{ch},k,t}-P_{\mathrm{dch},k,t}/e_{\mathrm{dch},k}\right)\Delta t/E_k\quad, \\ \forall k\in\mathcal{G}_{\mathrm{ES}},t\in\mathcal{T} \tag{4d}$$

$$soc_{k,\min}\le soc_{k,t}\le soc_{k,\max}\quad,\quad\forall k\in\mathcal{G}_{\mathrm{ES}},t\in\mathcal{T} \tag{4e}$$

$$\alpha_{\mathrm{ch},k,t}+\alpha_{\mathrm{dch},k,t}\le 1\quad,\quad 0\le P_{\mathrm{ch},k,t}\le\alpha_{\mathrm{ch},k,t}P_{\mathrm{ch},k,\max}\quad, \\ 0\le P_{\mathrm{dch},k,t}\le\alpha_{\mathrm{dch},k,t}P_{\mathrm{dch},k,\max}\quad,-Q_{k,\max}\le Q_{k,t}\le Q_{k,\max}\;, \\ P_{k,t}=P_{\mathrm{dch},k,t}-P_{\mathrm{ch},k,t}\quad,\quad\forall k\in\mathcal{G}_{\mathrm{ES}},t\in\mathcal{T} \tag{4f}$$

$$P_{k,t}^2+Q_{k,t}^2\le S_{k,\mathrm{app}}^2\quad,\quad\forall k\in\mathcal{G}_{\mathrm{ES}},t\in\mathcal{T} \tag{4g}$$

### *3.3. Modeling of Integrated Power and Gas DRs*

The dispersed customers participating in DR often have intricately intertwined power and gas supplies. For example, customers receiving power from the same power node (PN) may receive gas from various gas nodes (GN). This poses challenges in determining the scale of DR modeling and dispatching. While the most direct and precise approach would involve dispatching power or gas DR for each individual customer, this may be impractical due to the resulting model's complexity and associated computational burden, especially considering the growing prevalence of DR and the urgency of decision-making in emergencies. Alternatively, if we take a step back and dispatch DR for individual PNs (or GNs), a common approach in existing related research, we may still encounter challenges. For example, if we dispatch power and gas DRs for a specific PN, *i.e.*, for all customers it serves, it allows for straightforward quantification of reduced power demand at that PN. However, mapping the gas demand reductions of these customers to specific GNs is challenging, as customers may have distinct gas supply sources. Additionally, dispatching power DR at the PN level and gas DR at the GN level separately may raise equity concerns. Customers located at the intersection of a PN and GN, if both are called for DR, may face a more pronounced cumulative reduction in both power and gas demands, compared to customers affected by only one form of DR, a challenge that is difficult to address in modeling.

Therefore, we propose a straightforward source-based zoning method that partitions customers participating in power or gas DR into distinct zones or groups, each supplied by the same PN and GN, as shown in Fig. 3. An incidence matrix is then created, where a “1” in the *i*-th row and *j*-th column indicates the presence of a group supplied by GN *i* and PN *j*, and a “0” indicates the absence of such a group.

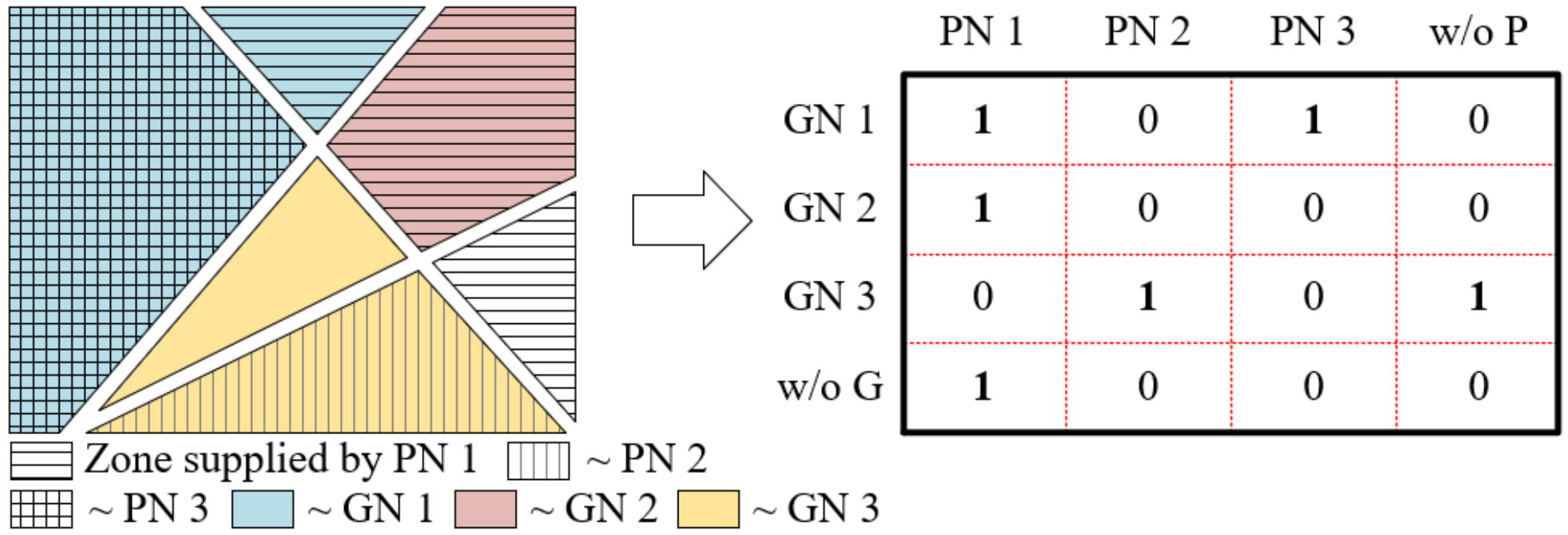

Fig. 3. The zoning method and the resulting incidence matrix.

Then, power and gas DRs can be dispatched for groups. When implemented, the dispatch decisions for each group can be further allocated to its customers, *e.g.*, proportionally based on their submitted DR capacities. Moreover, the dispersed energy demand reductions among the groups can be easily mapped to specific PNs and GNs of the IENDS. For example, the gas demand reduction at GN *i* can be derived by summing the reductions from gas DRs in the zones marked by “1” in the *i*-th row of the incidence matrix. Additionally, power and gas DRs within the same group should be mutually restricted to prevent excessive reduction in overall energy demand, which could endanger health, safety, and essential business and factory operations. Therefore, the dispatching of integrated power and gas DRs is modeled by (5), with power DR as an example. The dispatchable form of DR, *i.e.*, DLC, is assumed.

$$\gamma_{\mathrm{P},(i,j),t} \le \delta_{\mathrm{P},j,t} \quad , \quad \forall (i,j) \in \mathcal{Z}_{\mathrm{DR}}, t \in \mathcal{T} \tag{5a}$$

$$\sum_{t\in\mathcal{T}} \gamma_{\mathrm{P},(i,j),t} \le TP_{(i,j),\max} \quad , \quad \forall (i,j) \in \mathcal{Z}_{\mathrm{DR}} \tag{5b}$$

$$\begin{gathered} \sum_{l=0}^{TP_{(i,j),\mathrm{du},\max}} \gamma_{\mathrm{P},(i,j),t+l} \le TP_{(i,j),\mathrm{du},\max} \quad , \\ \forall (i,j) \in \mathcal{Z}_{\mathrm{DR}}, t \in \left\{ t \mid t \le D - TP_{(i,j),\mathrm{du},\max}, t \in \mathcal{T} \right\} \end{gathered} \tag{5c}$$

$$\begin{gathered} \sum_{l=0}^{TP_{(i,j),\mathrm{du},\min}-1} \gamma_{\mathrm{P},(i,j),t+l} \ge \left( \gamma_{\mathrm{P},(i,j),t} - \gamma_{\mathrm{P},(i,j),t-1} \right) TP_{(i,j),\mathrm{du},\min} \quad , \\ \forall (i,j) \in \mathcal{Z}_{\mathrm{DR}}, t \in \left\{ t \mid t \le D - TP_{(i,j),\mathrm{du},\min} + 1, t \in \mathcal{T} \right\} \end{gathered} \tag{5d}$$

$$\begin{gathered} \sum_{l=0}^{TP_{(i,j),\mathrm{int},\min}-1} \left( 1 - \gamma_{\mathrm{P},(i,j),t+l} \right) \ge \left( \gamma_{\mathrm{P},(i,j),t-1} - \gamma_{\mathrm{P},(i,j),t} \right) TP_{(i,j),\mathrm{int},\min} \quad , \\ \forall (i,j) \in \mathcal{Z}_{\mathrm{DR}}, t \in \left\{ t \mid t \le D - TP_{(i,j),\mathrm{int},\min} + 1, t \in \mathcal{T} \right\} \end{gathered} \tag{5e}$$

$$\begin{gathered} \sigma_{\mathrm{P},(i,j),\min} \tilde{P}_{(i,j),t} \gamma_{\mathrm{P},(i,j),t} \le P_{\mathrm{DR},(i,j),t} \le \sigma_{\mathrm{P},(i,j),\max} \tilde{P}_{(i,j),t} \gamma_{\mathrm{P},(i,j),t} \quad , \\ Q_{\mathrm{DR},(i,j),t} = \varphi_{(i,j)} P_{\mathrm{DR},(i,j),t} \quad , \forall (i,j) \in \mathcal{Z}_{\mathrm{DR}}, t \in \mathcal{T} \end{gathered} \tag{5f}$$

$$\begin{gathered} \sigma_{\mathrm{N},(i,j),\min} \tilde{F}_{(i,j),t} \gamma_{\mathrm{N},(i,j),t} \le F_{\mathrm{DR},(i,j),t} \le \sigma_{\mathrm{N},(i,j),\max} \tilde{F}_{(i,j),t} \gamma_{\mathrm{N},(i,j),t} \; , \\ \forall (i,j) \in \mathcal{Z}_{\mathrm{DR}}, t \in \mathcal{T} \end{gathered} \tag{5g}$$

$$\frac{P_{\mathrm{DR},(i,j),t}}{\tilde{P}_{(i,j),t}}+\frac{F_{\mathrm{DR},(i,j),t}}{\tilde{F}_{(i,j),t}} \le H_{(i,j)} \quad , \quad \forall (i,j) \in \mathcal{Z}_{\mathrm{DR}}, t \in \mathcal{T} \tag{5h}$$

$$P_{\mathrm{DR},j,t} = \sum_{i:(i,j)\in\mathcal{Z}_{\mathrm{DR}}} P_{\mathrm{DR},(i,j),t} \quad , \quad Q_{\mathrm{DR},j,t} = \sum_{i:(i,j)\in\mathcal{Z}_{\mathrm{DR}}} Q_{\mathrm{DR},(i,j),t} \quad , \quad \forall j \in \mathcal{N}_{\mathrm{PN.DR}}, t \in \mathcal{T} \tag{5i}$$

$$F_{\mathrm{DR},i,t} = \sum_{j:(i,j)\in\mathcal{Z}_{\mathrm{DR}}} F_{\mathrm{DR},(i,j),t} \quad , \quad \forall i \in \mathcal{N}_{\mathrm{GN.DR}}, t \in \mathcal{T} \tag{5j}$$

Constraint (5a) states that power DR can be implemented as long as the supplying PN is energized. Constraints (5b)-(5e) define the limits of a DR event in terms of total duration, duration per event, and the interval between two consecutive events. Similar constraints for gas DR can be formulated. Then, constraints (5f) and (5g) bound the power and gas demand reductions due to DR, respectively, while (5h) sets an upper limit on the combined power and gas demand reduction. For a more aggressive strategy, *e.g.*, where power and gas DRs cannot be implemented simultaneously, (5h) can be modified to "$\gamma_{\mathrm{P},(i,j),t}+\gamma_{\mathrm{N},(i,j),t}\le 1$". Constraints (5i) and (5j) map power and gas demand reductions among groups to specific PNs and GNs, respectively, based on the incidence matrix.

*3.4. Modeling of Repair activities*

We consider an RU to be a combination of crews and equipment that can move and independently complete a repair task, and assume multiple identical RUs are available for the IENDS. The modeling of repair process incorporates several factors to better simulate real-world conditions: 1) Once an RU arrives to repair a facility, it continues working until the facility is fully repaired. 2) Repair efficiency varies with the number of active RUs, and additional RUs can join an ongoing repair to accelerate it. Specifically, 1) can be expressed by:

$$x_{j,i,t-1} - x_{j,i,t} \le \kappa_{i,t} \quad , \quad \forall j \in \mathcal{M}_{\mathrm{RU}}, i \in \mathcal{N}_{\mathrm{DMG}}, t \in \mathcal{T} \tag{6a}$$

We assign a health level to each damaged facility to indicate its state, ranging from 0 (no repair) to 1 (full repair). The repair process thus involves restoring the health level to 1. Therefore, factor 2) can be formulated by (6b). $\eta_{i,t}$ represents the actual repair efficiency, while $\beta_{i,y}$ is the efficiency under the condition of $y$ working RUs and can be simply estimated.

$$\sum_{j\in\mathcal{M}_{\mathrm{RU}}} x_{j,i,t} = \sum_{y=0}^{|\mathcal{M}_{\mathrm{RU}}|} y b_{i,t,y} \quad , \quad \eta_{i,t} = \sum_{y=0}^{|\mathcal{M}_{\mathrm{RU}}|} \beta_{i,y} b_{i,t,y} \quad , \quad \sum_{y=0}^{|\mathcal{M}_{\mathrm{RU}}|} b_{i,t,y} = 1 \quad , \quad \forall i \in \mathcal{N}_{\mathrm{DMG}}, t \in \mathcal{T} \tag{6b}$$

Furthermore, a damaged facility can return to normal once its health level rises to 1, as indicated by (6c) and (6d).

$$\kappa_{i,t} \le \sum_{\tau=1}^{t-1} \eta_{i,\tau} \quad , \quad \kappa_{i,t-1} \le \kappa_{i,t} \quad , \forall i \in \mathcal{N}_{\mathrm{DMG}}, t \in \mathcal{T} \setminus \{1\} \tag{6c}$$

$$\kappa_{i,1} = 0 \quad , \quad \forall i \in \mathcal{N}_{\mathrm{DMG}} \tag{6d}$$

*3.5. Modeling of Operation for the EPDS Component*

The operation of the EPDS component can be expressed as follows based on the linearized DistFlow model [38]:

$$\sum_{(i',i)\in\mathcal{B}} P_{i'i,t} + \sum_{k\in\mathcal{G}(i)} P_{k,t} = \sum_{(i,i')\in\mathcal{B}} P_{ii',t} + \delta_{\mathrm{P},i,t}\tilde{P}_{i,t} - [\![ i \in \mathcal{N}_{\mathrm{PN.DR}} ]\!] P_{\mathrm{DR},i,t} + \sum_{k\in\mathcal{C}(i)} P_{\mathrm{com},k,t} + \sum_{k\in\mathcal{S}(i)} P_{\mathrm{stg},k,t} \quad , \forall i \in \mathcal{N}_{\mathrm{PN}}, t \in \mathcal{T} \tag{7a}$$

$$\sum_{(i',i)\in\mathcal{B}} Q_{i'i,t} + \sum_{k\in\mathcal{G}(i)} Q_{k,t} = \sum_{(i,i')\in\mathcal{B}} Q_{ii',t} + \delta_{\mathrm{P},i,t}\tilde{Q}_{i,t} - [\![ i \in \mathcal{N}_{\mathrm{PN.DR}} ]\!] Q_{\mathrm{DR},i,t} + \sum_{k\in\mathcal{C}(i)} \varphi_k P_{\mathrm{com},k,t} + \sum_{k\in\mathcal{S}(i)} \varphi_k P_{\mathrm{stg},k,t} \quad , \forall i \in \mathcal{N}_{\mathrm{PN}}, t \in \mathcal{T} \tag{7b}$$

$$
\begin{aligned}
& u_{i',t}^2 \geq u_{i,t}^2 - 2\left(P_{ii',t} r_{ii'} + Q_{ii',t} x_{ii'}\right) - \left[\!\left[ sno(i,i') \in \mathcal{N}_{\mathrm{DMG}} \right]\!\right] M\left(1-\kappa_{sno(i,i'),t}\right) \quad , \\
& u_{i',t}^2 \leq u_{i,t}^2 - 2\left(P_{ii',t} r_{ii'} + Q_{ii',t} x_{ii'}\right) + \left[\!\left[ sno(i,i') \in \mathcal{N}_{\mathrm{DMG}} \right]\!\right] M\left(1-\kappa_{sno(i,i'),t}\right) \quad , \\
& \forall (i,i') \in \mathcal{B}, t \in \mathcal{T}
\end{aligned} \tag{7c}
$$

$$
u_{i,\min}^2 \leq u_{i,t}^2 \leq u_{i,\max}^2 \quad , \quad \forall i \in \mathcal{N}_{\mathrm{PN}}, t \in \mathcal{T} \tag{7d}
$$

$$
\begin{aligned}
& P_{ii',t}^2 + Q_{ii',t}^2 \leq \left[\left(\kappa_{sno(i,i'),t} - 1\right)\left[\!\left[ sno(i,i') \in \mathcal{N}_{\mathrm{DMG}} \right]\!\right] + 1\right] S_{(i,i'),\mathrm{app}}^2 , \\
& \forall (i,i') \in \mathcal{B}, t \in \mathcal{T}
\end{aligned} \tag{7e}
$$

$$
\delta_{\mathrm{P},i,t-1} \leq \delta_{\mathrm{P},i,t} \quad , \quad \forall i \in \mathcal{N}_{\mathrm{PN}}, t \in \mathcal{T} \setminus \{1\} \tag{7f}
$$

Specifically, the Iverson bracket $[\![\cdot]\!]$ takes a value of 1 if the inner condition is satisfied and 0 otherwise. Constraint (7e) can be linearized by the method in [38]. $sno(i, i')$ represents the series number of branch $(i, i')$, and $[\![sno(i, i') \in \mathcal{N}_{\mathrm{DMG}}]\!]$ indicates whether branch $(i, i')$ is damaged. Constraint (7f) ensures that the restored load cannot be de-energized.

### *3.6. Modeling of Operation for the NGDS Component*

#### *3.6.1. Operational constraints*

The operation of NGDS can be formulated by:

$$
\begin{aligned}
& \sum_{(i,'i) \in \mathcal{P}} F_{i'i,t} + \sum_{k \in \mathcal{W}(i)} L_{\mathrm{src},k,t} + \sum_{k \in \mathcal{N}_{\mathrm{GAS}}^{\mathrm{r}}(i)} L_{\mathrm{stg},k,t} = \sum_{(i,i') \in \mathcal{P}} F_{ii',t} + \\
& \delta_{\mathrm{N},i,t} \tilde{F}_{i,t} - \left[\!\left[ i \in \mathcal{N}_{\mathrm{GN.DR}} \right]\!\right] F_{\mathrm{DR},i,t} + \sum_{k \in \mathcal{F}(i)} L_{k,t} \quad , \forall i \in \mathcal{N}_{\mathrm{GN}}, t \in \mathcal{T}
\end{aligned} \tag{8a}
$$

$$
\pi_{i,t}^2 - \pi_{i',t}^2 = \mathrm{sgn}\left(F_{ii',t}\right) K_{ii'} F_{ii',t}^2 \quad , \quad \forall (i,i') \in \mathcal{P}_{\mathrm{pa}}, t \in \mathcal{T} \tag{8b}
$$

$$
0 \leq F_{k_1 k_2,t} \leq F_{k_1 k_2,\max} \quad , \quad \forall k \in \mathcal{P}_{\mathrm{com}}, t \in \mathcal{T} \tag{8c}
$$

$$
\begin{aligned}
& \zeta_{k,\min}^2 \pi_{k_1,t}^2 - \pi_{k_2,t}^2 \leq M\left(1-\Lambda_{k,t}\right) \quad , \quad \pi_{k_2,t}^2 - \zeta_{k,\max}^2 \pi_{k_1,t}^2 \leq M\left(1-\Lambda_{k,t}\right) \quad , \\
& \forall k \in \mathcal{P}_{\mathrm{com}}, t \in \mathcal{T}
\end{aligned} \tag{8d}
$$

$$
\begin{aligned}
& -M\left(1-\Lambda_{k,t}\right) \leq P_{\mathrm{com},k,t} - \rho_{\mathrm{com},k} F_{k_1 k_2,t} \leq M\left(1-\Lambda_{k,t}\right) \quad , \\
& \forall k \in \mathcal{P}_{\mathrm{com}}, t \in \mathcal{T}
\end{aligned} \tag{8e}
$$

$$
-M\Lambda_{k,t} \leq \pi_{k_1,t}^2 - \pi_{k_2,t}^2 \leq M\Lambda_{k,t} \quad , \quad \forall k \in \mathcal{P}_{\mathrm{com}}, t \in \mathcal{T} \tag{8f}
$$

$$
0 \leq P_{\mathrm{com},k,t} \leq M\Lambda_{k,t} \quad , \quad \forall k \in \mathcal{P}_{\mathrm{com}}, t \in \mathcal{T} \tag{8g}
$$

$$
P_{\mathrm{stg},i,t} = \rho_{\mathrm{stg,inj},i} L_{\mathrm{inj},i,t} + \rho_{\mathrm{stg,wd},i} L_{\mathrm{wd},i,t} , \forall i \in \mathcal{N}_{\mathrm{GAS}}^{\mathrm{r}}, t \in \mathcal{T} \tag{8h}
$$

$$
\pi_{i,\min}^2 \leq \pi_{i,t}^2 \leq \pi_{i,\max}^2 \quad , \quad \forall i \in \mathcal{N}_{\mathrm{GN}}, t \in \mathcal{T} \tag{8i}
$$

$$
L_{\mathrm{src},k,\min} \leq L_{\mathrm{src},k,t} \leq L_{\mathrm{src},k,\max} \quad , \quad \forall k \in \mathcal{W}, t \in \mathcal{T} \tag{8j}
$$

$$
\delta_{\mathrm{N},i,t-1} \leq \delta_{\mathrm{N},i,t} \quad , \quad \forall i \in \mathcal{N}_{\mathrm{GN}}, t \in \mathcal{T} \setminus \{1\} \tag{8k}
$$

Constraint (8a) ensures gas flow balance. Constraint (8b) represents the Weymouth equation, governing the relationship between gas flow and pressure drop along passive pipelines without compressors. Constraints (8c) to (8g) describe the operation of EDCs. $k_1$ and $k_2$ represent the GNs at the suction and discharge sides of EDC $k$. The out-of-service situation of EDC when losing power is expressed by (8f) and (8g), where gas is considered to be bypassed with equal suction and discharge pressures. Constraint (8h) denotes the power consumption of gas storages. Constraint (8i) bounds the pressure of GNs, (8j) limits gas injection from sources, and (8k) prevents the restored load from being interrupted again.

### 3.6.2. Convexified Relaxation of the Weymouth Equation

The presence of sgn(*) and nonconvexity of constraint (8b) hinder the solvability of the whole programming model. To address this, the following procedure is implemented to realize a second-order cone relaxation. First, we rewrite (8b) as (9a) and (9b) to release sgn(*):

$$K_{ii'}F_{ii',t}^2 = Y_{ii',t} \tag{9a}$$

$$\left[\!\left[ F_{ii',t} \geq 0 \ , \ Y_{ii',t} = \pi_{i,t}^2 - \pi_{i',t}^2 \right]\!\right] + \left[\!\left[ F_{ii',t} < 0 \ , \ Y_{ii',t} = \pi_{i',t}^2 - \pi_{i,t}^2 \right]\!\right] = 1 \tag{9b}$$

The logical expression in (9b) can be further formulated by linear constraints (9c) and (9d) based on the big-M methods, in which the bidirectional gas flow is additionally bounded within the proper range:

$$\begin{aligned} & -F_{ii',t} \leq F_{ii',\max}\left(1 - y_{ii',t}\right) \ , \\ & -M\left(1 - y_{ii',t}\right) \leq Y_{ii',t} - \left(\pi_{i,t}^2 - \pi_{i',t}^2\right) \leq M\left(1 - y_{ii',t}\right) \end{aligned} \tag{9c}$$

$$\begin{aligned} & F_{ii',t} \leq F_{ii',\max} y_{ii',t} \ , \\ & -My_{ii',t} \leq Y_{ii',t} - \left(\pi_{i',t}^2 - \pi_{i,t}^2\right) \leq My_{ii',t} \end{aligned} \tag{9d}$$

Then, (9a) is relaxed to (9e), which can be further rewritten as the second-order cone form in (9f):

$$K_{ii'}F_{ii',t}^2 \leq Y_{ii',t} \tag{9e}$$

$$\left\| \begin{matrix} 2\sqrt{K_{ii'}}F_{ii',t} \\ Y_{ii',t} - 1 \end{matrix} \right\|_2 \leq Y_{ii',t} + 1 \tag{9f}$$

The model becomes tractable following the above conversion. To guarantee the exactness of the relaxation, a simple but effective penalty term will be added to the objective function.

### 3.7. Modeling of Objective Function

The proposed strategy aims to maximize the restoration performance of the IENDS, which can be formulated by:

$$\begin{aligned} \max \ & \underbrace{\varsigma_1 \frac{\sum_{t\in\mathcal{T}}\sum_{i\in\mathcal{N}_{\mathrm{PN}}} w_{\mathrm{P},i}\left(\delta_{\mathrm{P},i}\tilde{P}_{i,t} - \left[\!\left[ i \in \mathcal{N}_{\mathrm{PN.DR}} \right]\!\right] P_{\mathrm{DR},i,t}\right)\Delta t}{\sum_{t\in\mathcal{T}}\sum_{i\in\mathcal{N}_{\mathrm{PN}}} w_i \tilde{P}_{i,t}\Delta t}}_{\mathrm{i}} + \underbrace{\varsigma_2 \frac{\sum_{t\in\mathcal{T}}\sum_{i\in\mathcal{N}_{\mathrm{GN}}} w_{\mathrm{N},i}\left(\delta_{\mathrm{N},i}\tilde{F}_{i,t} - \left[\!\left[ i \in \mathcal{N}_{\mathrm{GN.DR}} \right]\!\right] F_{\mathrm{DR},i,t}\right)\Delta t}{\sum_{t\in\mathcal{T}}\sum_{i\in\mathcal{N}_{\mathrm{GN}}} w_i \tilde{F}_{i,t}\Delta t}}_{\mathrm{ii}} - \\ & \underbrace{o_1 \sum_{t\in\mathcal{T}}\sum_{(i,i')\in\mathcal{P}_{\mathrm{pa}}} Y_{ii',t}}_{\mathrm{iii}} - \underbrace{o_2 \sum_{t\in\mathcal{T}}\left( \sum_{j\in\mathcal{M}_{\mathrm{RU}}}\sum_{i\in\mathcal{N}_{\mathrm{DMG}}} v_{j,i,t} + \sum_{j\in\mathcal{M}_{\mathrm{GAS}}}\sum_{i\in\mathcal{N}_{\mathrm{GAS}}} v_{j,i,t} + \sum_{j\in\mathcal{M}_{\mathrm{DSL}}}\sum_{i\in\mathcal{N}_{\mathrm{DSL}}} v_{j,i,t} \right)}_{\mathrm{iv}} - \\ & \underbrace{o_3 \sum_{t\in\mathcal{T}}\left( \sum_{j\in\mathcal{M}_{\mathrm{GAS}}}\sum_{i\in\mathcal{N}_{\mathrm{GAS}}} \chi_{j,i,t} + \sum_{j\in\mathcal{M}_{\mathrm{DSL}}}\sum_{i\in\mathcal{N}_{\mathrm{DSL}}} \chi_{j,i,t} \right)}_{\mathrm{v}} - \underbrace{o_4 \sum_{t\in\mathcal{T}}\left( \sum_{k\in\mathcal{G}_{\mathrm{D}}\cup\mathcal{G}_{\mathrm{dual}}} C_{\mathrm{D},k,t} + \sum_{k\in\mathcal{G}_{\mathrm{N}}\cup\mathcal{G}_{\mathrm{dual}}} C_{\mathrm{N},k,t} \right)}_{\mathrm{vi}} \end{aligned} \tag{10}$$

Specifically, terms i and ii represent the total supplied electricity and gas demands, respectively. Term iii serves as a penalty to tighten the relaxation of the Weymouth equation, which has proven exact [20], [39]. Term iv denotes the cost of deploying mobile resources. Terms iv and v are penalties to prevent futile movement and fuel exchanges. Term vi represents the cost of generators providing emergency power and also prevents superfluous generation after reconnection to the bulk grid. The metric in (10) is a commonly used resilience quantification indicator, where the first two terms can be interpreted as the time integral of system performance (indicated by the supplied load) during the scheduling horizon, while the remaining terms accounts for costs associated with restoration efforts [40].

## 4. Case Studies

Case studies are conducted in this section. The constructed model is coded in MATLAB R2020b with YALMIP toolbox [41] and solved using Gurobi v10.0.1 on a computer with an Intel Core i7-11800H CPU and 16 GB RAM.

### *4.1. Test System and Scenario*

The test IENDS comprises a modified IEEE 37-node EPDS and an 8-node NGDS, as shown in Fig. 4(a). The details of the 37-node EPDS can be found in [42], and the main parameters of the 8-node NGDS are given in Table 1. Load curves in Fig. 5 are based on actual winter load data from El Paso, Texas, USA [43] and serve as multipliers to base loads to simulate demand variations. 2 DFUs, 1 NGFU, and 1 dual-fuel unit are involved. The NGFU and dual-fuel unit are supplied from GN 6 and GN 4, respectively. Based on subsection III.C, four zones are assumed to participate in power and gas DR, with the incidence matrix provided in Fig. 4(b). The energy storage system is installed at PN 30, and gas storage at GN 8. Meanwhile, a diesel reservoir is located in the area. The main parameters of these facilities are primarily based on real-world data and are summarized in Table 1.

Assume the IENDS has six overhead branches damaged by a disaster. Moreover, we limit the gas input to the NGDS to 1800 $Sm^3/h$, which is below the total gas demand in some periods, to simulate upstream gas shortages commonly seen in past events [5], [9]. 2 diesel tankers and 2 gas tankers are dispatched. All tankers and onsite fuel storages are initially empty. 3 RUs are available to repair damaged branches, with efficiencies given in Table 1. These mobile resources are assumed to have the same travel time between any two given locations, set arbitrarily to 1 h or 2 h for simplicity. Suppose the restoration initiates at 4:00, a crunch time with a surge in energy demand, particularly for the natural gas demand. Next, we will examine how the IENDS will be restored over the next 12 hours, with a time step $\Delta t$ of 1 h. For simplicity, we set $\varsigma_{1/2}$ and $o_{1/2/3/4}$ in the objective function to 1 and $10^{-3}$, respectively.

Table 1. Parameter settings of the test systems

<table>
<tr><td colspan="2"><b>Generators</b></td><td colspan="2">DFU</td><td colspan="2">NGFU</td><td colspan="2">Dual-fuel unit</td></tr>
<tr><td colspan="2">Rated power (kW)/capacity (kVA)</td><td colspan="2">450/562.5</td><td colspan="2">500/625</td><td colspan="2">450/562.5 (diesel),<br>400/500 (gas)</td></tr>
<tr><td colspan="2">Fuel consumption at 1/4, 1/2, 3/4, and full load (L/h or $Sm^3/h$)</td><td colspan="2">45, 79, 106, 132</td><td colspan="2">69, 109, 153, 198</td><td colspan="2">45, 79, 106, 132 (diesel),<br>70, 99, 132, 168 (gas)</td></tr>
<tr><td colspan="2"><b>Fuel storages</b></td><td colspan="2">Diesel reservoir</td><td colspan="2">Diesel/gas tankers</td><td colspan="2">Onsite diesel/gas storage</td></tr>
<tr><td colspan="2">Capacity (L or $Sm^3$)</td><td colspan="2">$1\times10^4$</td><td colspan="2">300/500</td><td colspan="2">1000/1000</td></tr>
<tr><td colspan="4"><b>DR (the same settings for power and gas DR)</b></td><td colspan="2"><b>Energy storage</b></td><td colspan="2"><b>Repair units</b></td></tr>
<tr><td>$TP_{(i,j),\max}$ (h)</td><td>6</td><td>$TP_{(i,j),\text{int},\min}$ (h)</td><td>3</td><td>$P_{\text{ch/dch},k,\max}$ (kW), $E_k$ (kWh)</td><td>300, 600</td><td rowspan="3">$\beta_{i,y}$ when y=0, 1, 2, 3</td><td rowspan="3">0, 0.25, 0.5, 1</td></tr>
<tr><td rowspan="2">$TP_{(i,j),\text{du},\max}$, $TP_{(i,j),\text{du},\min}$ (h)</td><td rowspan="2">3, 1</td><td rowspan="2">$\sigma_{\text{P},(i,j),\min}$, $\sigma_{\text{P},(i,j),\max}$, $H_{(i,j)}$</td><td rowspan="2">0.2, 0.8, 1</td><td>$soc_{k,\min}$, $soc_{k,\max}$</td><td>0.1, 0.9</td></tr>
<tr><td>$e_{\text{ch/dch},k}$</td><td>0.95</td></tr>
<tr><td colspan="4"><b>Load in NGDS</b></td><td colspan="4"><b>Pipelines & compressors & gas storage</b></td></tr>
<tr><td>Node</td><td>Load ($Sm^3/h$)</td><td>Node</td><td>Load ($Sm^3/h$)</td><td colspan="2">$K_{ii'}$ [kPa/($Sm^3/h$)]$^2$</td><td colspan="2">$3.7056\times10^{-5}$</td></tr>
<tr><td>1</td><td>0</td><td>5</td><td>350</td><td colspan="2">$\rho_{\text{com},k}$ [MW/($Sm^3/h$)]</td><td colspan="2">$4.2\times10^{-4}$</td></tr>
<tr><td>2</td><td>100</td><td>6</td><td>660</td><td colspan="2">$\zeta_{k,\min}$~$\zeta_{k,\max}$</td><td colspan="2">1.1~1.5</td></tr>
<tr><td>3</td><td>170</td><td>7</td><td>150</td><td colspan="2">$\psi_{i,max}$ for Gas storage ($Sm^3$)</td><td colspan="2">$1\times10^4$</td></tr>
<tr><td>4</td><td>470</td><td>8</td><td>150</td><td colspan="2">$e_{\text{inj},i}/e_{\text{wtd},i}$</td><td colspan="2">0.85</td></tr>
</table>

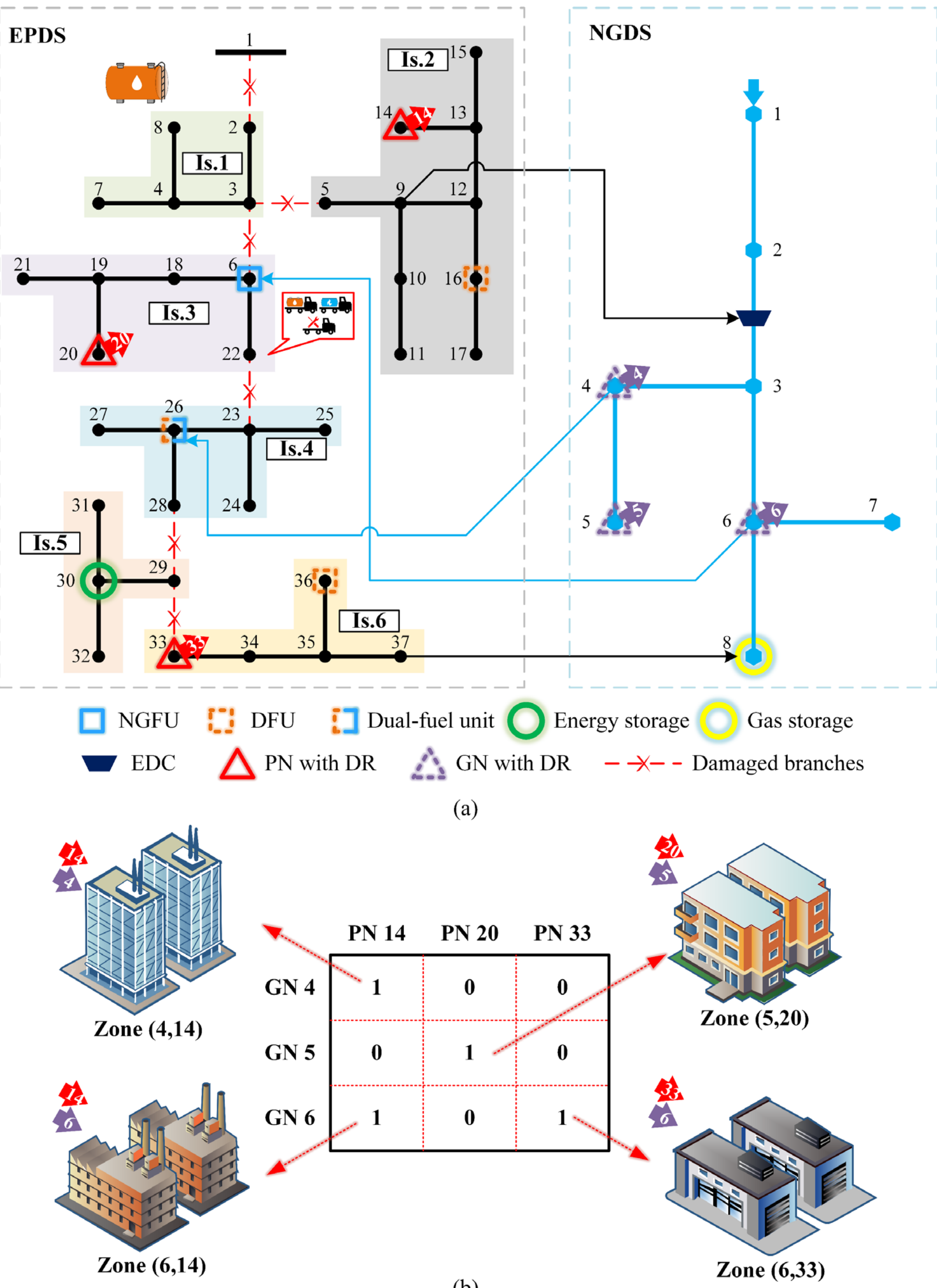

Fig. 4. (a) The test IENDS; (b) zones for power and gas DRs.

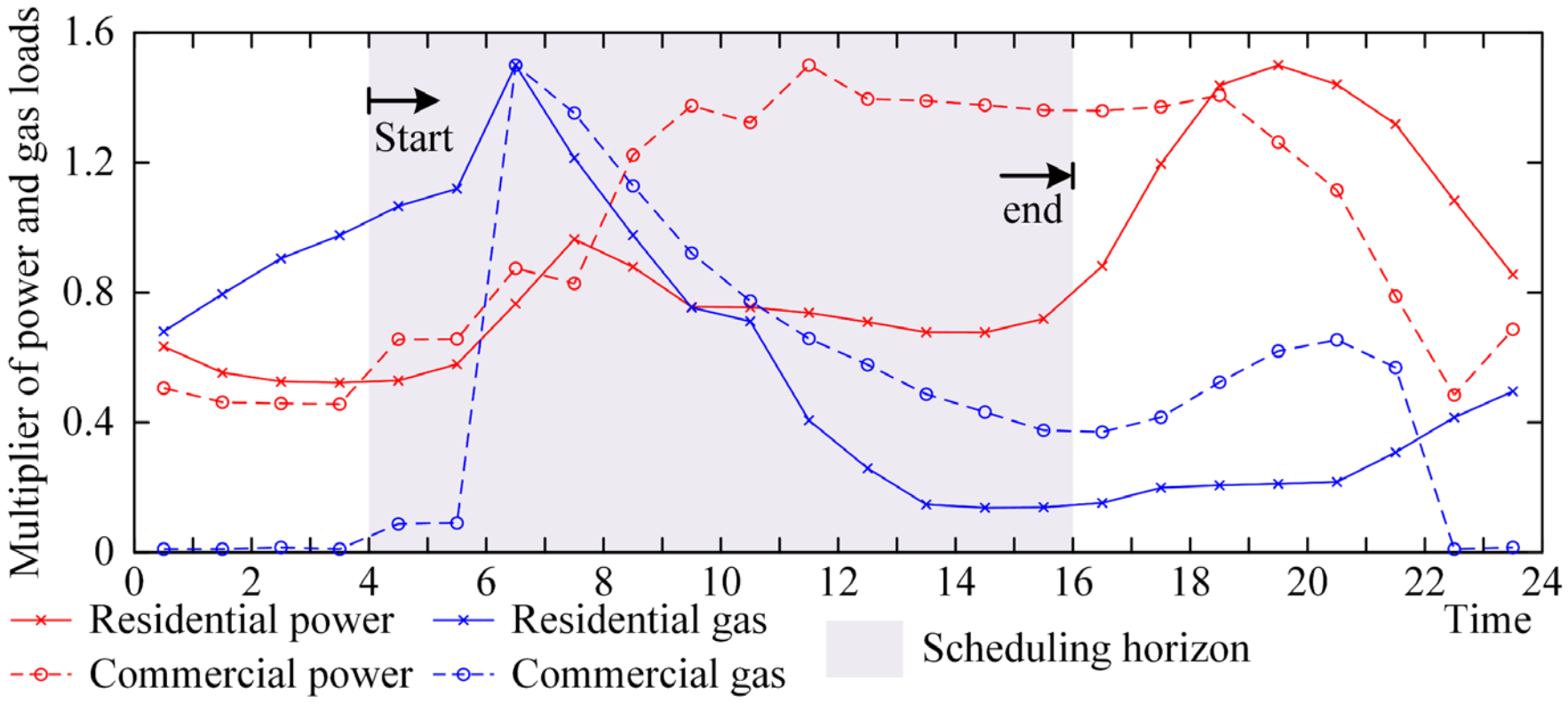

Fig. 5. The hourly power and gas load curves.

*4.2. Scheduling Results and Analysis*

The main results are in Figs. 6-8. The damaged branches initially create six islands. We will explore what happens on each island and how various resources are coordinated.

At the beginning, all RUs move to repair the damaged branch 1-2. Through their collaborative efforts with a high efficiency, branch 1-2 is fully repaired after 1 h, reintegrating **Is. 1** into the main grid and restoring its power loads.

The DFU in **Is. 2** is unavailable due to fuel shortage for the first three time spans (TSs) until diesel tanker 1, which has refueled at the reservoir, delivers diesel after the 3rd TS. Subsequently, the DFU operates with ample fuel to supply the local loads in **Is. 2** until it reconnects to the main grid when branch 3-5 is repaired at the end of the 4th TS. No power DR occurs in **Is. 2** due to sufficient power from the DFU.

**Is. 3** is supplied by the NGFU until it reconnects to the main grid following the repair of branch 3-6. Power DR events occur in **Is. 3** during the 3rd and 8th TSs, as shown in Fig. 8. The reasons are explained as follows. During the 3rd TS, there is a significant rise in gas demand; meanwhile, **Is. 2** has no available power, rendering the EDC unavailable to boost the pressure. As a result, the pressures of GNs 7 and 8 drop to their lower limits during the 3rd and 4th TSs. The NGDS falls into a stressed condition and cannot support the transmission of more gas. Consequently, there is a widespread call for intensive execution of gas DR in all zones during the 3rd and 4th TSs, and some of the GNs are not even picked up. Additionally, gas supply to the NGFU is limited by this stressed condition, leading to a power shortage in **Is. 3** and execution of power DR. In this period, the gas storage receives no power from **Is. 6** to support the NGDS. To alleviate the fuel stress on the NGFU, gas tanker 1 brings fuel during the 4th TS. The NGFU predominantly runs on onsite gas to supply **Is. 3** from the 4th to 7th TSs, and resumes NGDS supply as gas demand decreases in subsequent periods. Power DR also occurs during the 8th TS, because the power demand of **Is. 3** would slightly rise beyond the NGFU's capacity without DR.

The dual-fuel unit in **Is. 4** runs in gas mode throughout the scheduling, fueled by NGDS and gas tankers. Its fuel delivery process is similar to the NGFU in **Is. 3**. Initially, the unit relies on gas from NGDS until gas tanker 2 brings the fuel. Then, it runs on the onsite gas to circumvent the stressed condition of NGDS until the 8th TS, after which NGDS resumes supplying it. It is noteworthy that the power demand of **Is. 4** is always below its generation capacity at PN 26. As will be discussed later, this spare capacity is also effectively utilized.

The energy storage initially supports part of the loads in **Is. 5** but nearly depletes its stored energy by the 6th TS. After the repair of branch 28-29, **Is. 5** and **Is. 4** merge into a larger island, allowing the spare capacity of the dual-fuel unit in **Is.4** to supply extra loads of **Is. 5**, consistent with the findings in [44]. Specifically, right after the merger, this spare capacity is used to preemptively charge the energy storage. This ensures additional power is available when the upcoming restoration of extra loads (PN 29) increases the power demand beyond the dual-fuel unit's capacity. After jointly repairing branch 28-29, the RUs are dispersed to the remaining damaged branches, which will be repaired after the 11th TS, restoring all loads.

For **Is. 6**, its DFU cannot operate until diesel tanker 2 brings diesel from the reservoir after the 4th TS. However, this diesel is just enough for the DFU to run from the 5th to 8th TSs. To prevent further fuel shortage, diesel tanker 1, refueled at the reservoir, brings extra fuel to PN 36 at the end of the 8th TS. Due to the limited capacity of the tanker, this additional fuel supports only 2 h of full-power DFU operation, and a fuel shortage warning arises again by the end of the 10th TS. Fortunately, diesel tanker 2 returns to PN 36 after the 10th TS, carrying diesel from PN 16. This final fuel supplement ensures continuous DFU operation until **Is. 6** reconnects to the main grid. Note that for this final supplement, diesel tanker 2 is scheduled to refuel at PN 16 rather than the reservoir due to the shorter travel time required, as well as the surplus fuel at PN 16. Power DR is executed at PN 33 from the 9th to 11th TS because more loads are restored in this period and the limited DFU's capacity necessitates power DR for balance.

As described earlier, the NGDS enters a stressed condition from the 3rd TS due to a rise in gas demand and limited transmission capability, lasting for 2 or 3 TSs until the demand significantly decreases. Consequently, GN 6, which has a high demand, remains unrestored until the 4th TS, and gas DRs are extensively executed. Meanwhile, gas tankers are deployed to supply the NGFU and dual-fuel unit, relieving them from the stressed NGDS and reducing the burden on the NGDS for their feed. The NGDS exits the stressed condition from the 5th TS. DFU 36 starts up during the 5th TS, energizing the gas storage at PN 37. This enables the gas storage to inject gas into the NGDS, providing extra supply to the loads. Moreover, as **Is. 2** reconnects to the grid, the EDC gains sufficient power to boost the pressure of GN 3, ensuring the NGDS can support adequate gas transmission. In the following period, NGDS stress is largely alleviated, and no further DR is required.

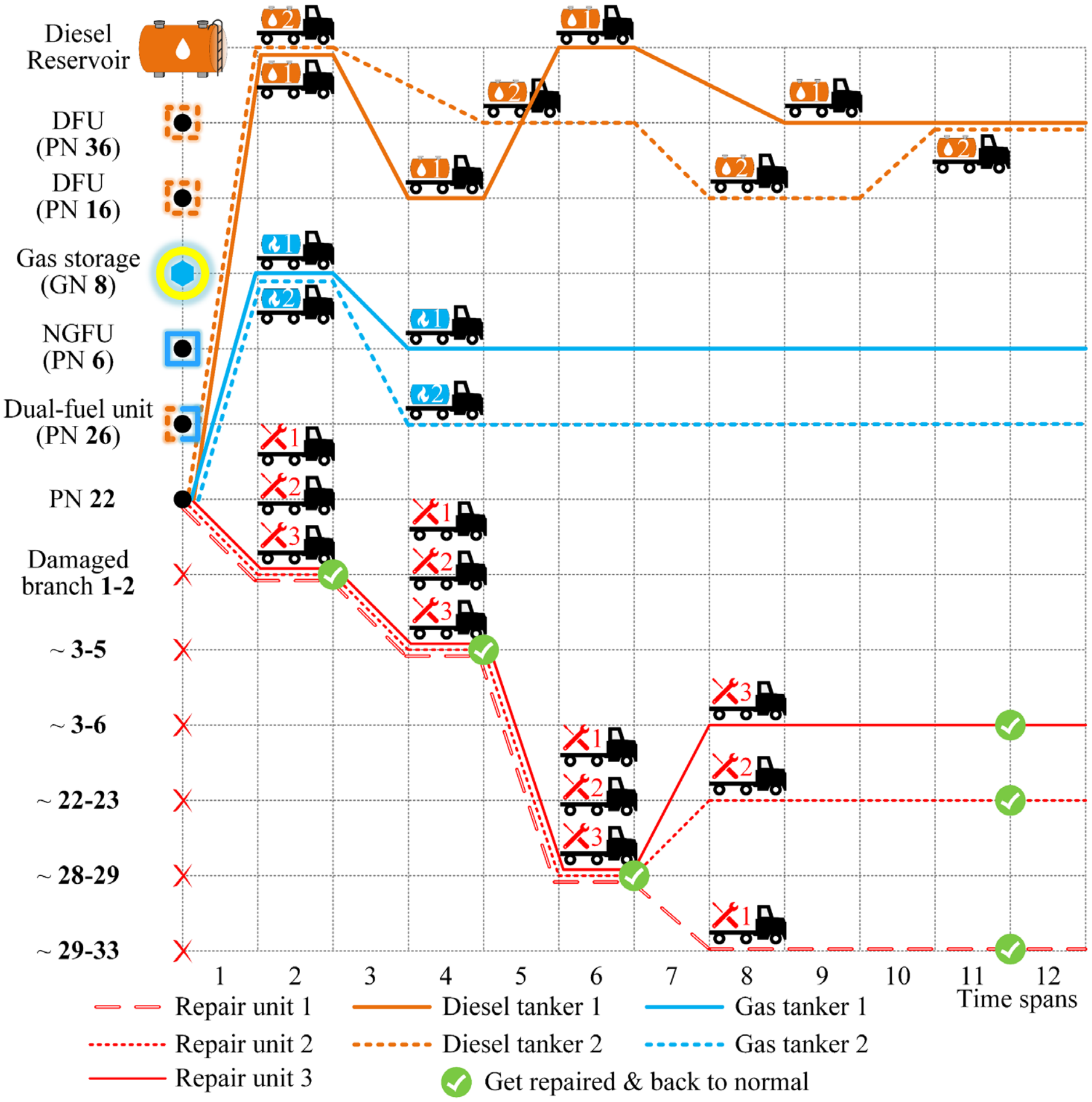

Fig. 6. Scheduling results of the mobile resources.

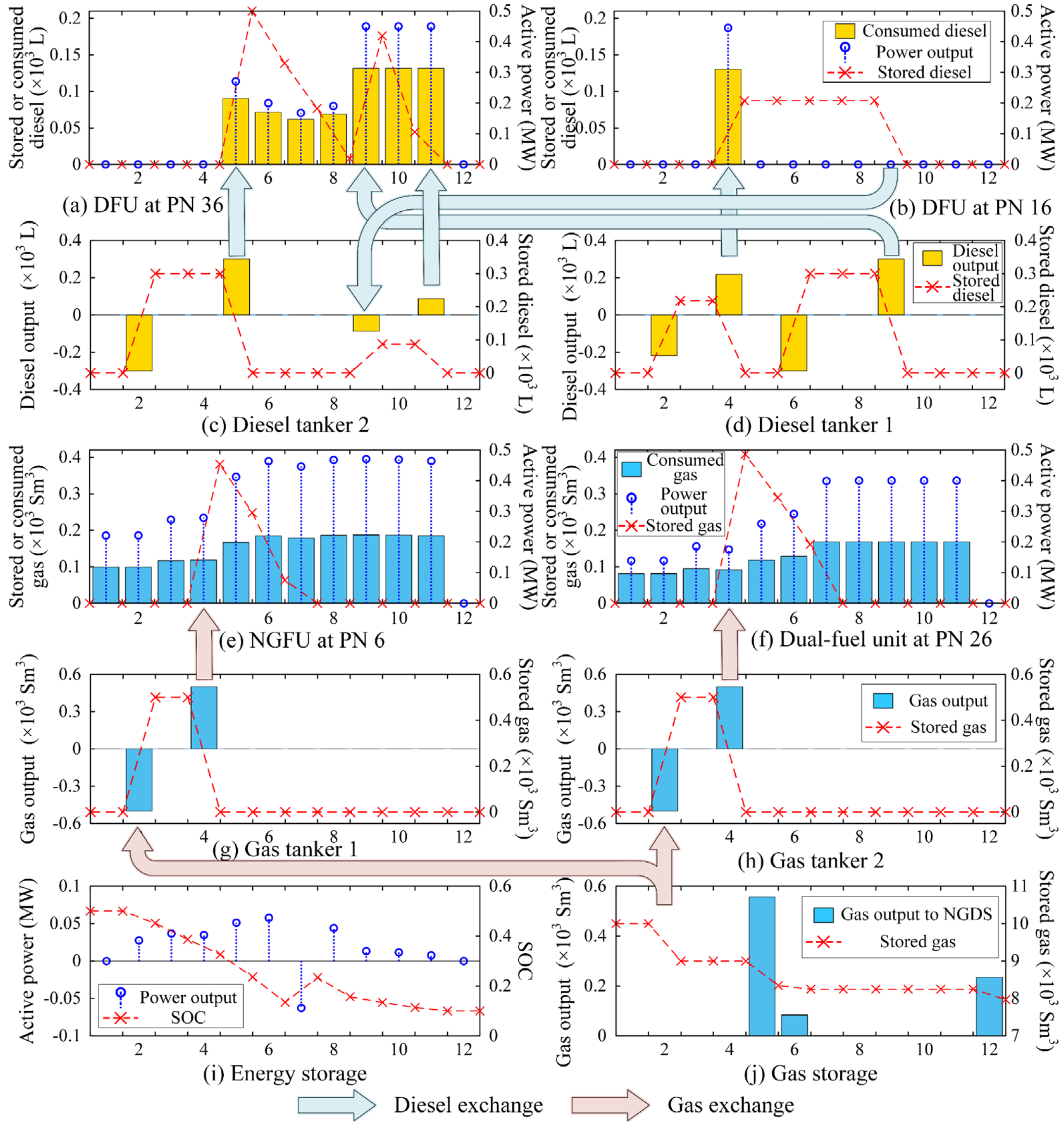

Fig. 7. Scheduling results of fuel delivery, with fuel transfer processes indicated by bold arrows. For example, as illustrated in (a) - (c), diesel tanker 2 supplies diesel to the (onsite storage of) DFU at PN 36 during the 5th time span, and refuels itself from PN 16 during the 9th time span.

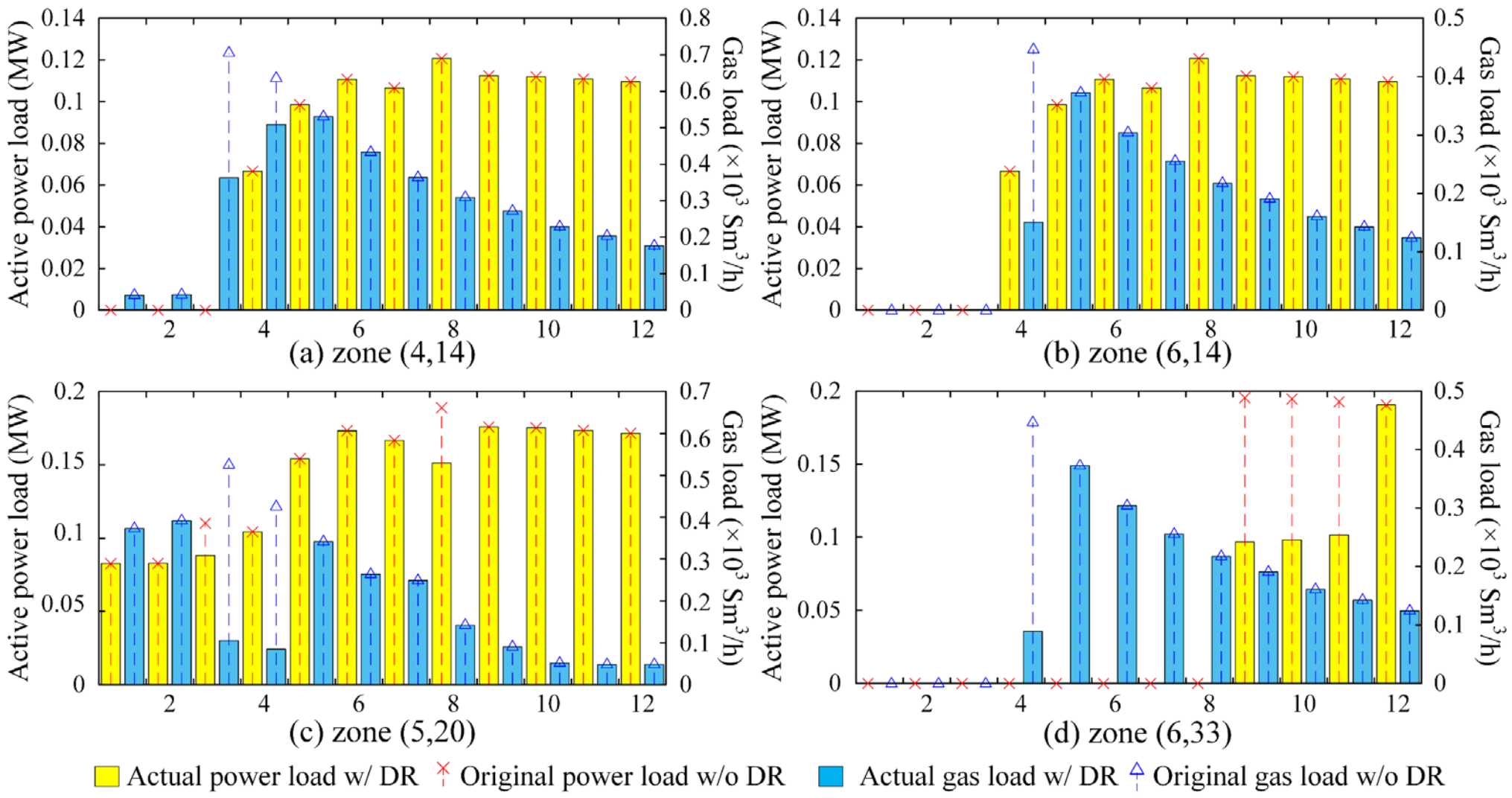

Fig. 8. Scheduling results of power and gas DRs.

*4.3. Comparison Among Scenarios*

The results demonstrate the effectiveness of the strategy. In addition to this **Base Case**, three other cases are explored:

**Case 1**: Cut off NGDS supply to NGFU and dual-fuel unit.

**Case 2**: Based on Case 1, allocate more fuel initially in all onsite storages, increasing from 0 to half of the capacity.

**Case 3**: Decrease the gas input to NGDS, from the maximum 1800 $Sm^3/h$ to 900 $Sm^3/h$, to simulate scenarios such as upstream pipeline freeze-offs.

For brevity, we focus on the key differences between each case and the **Base Case**. Fig. 9 presents the results of the hourly supplied demand, with the Y-axis calculated from the hourly component of the sum of the first two terms in (10).

The results of **Case 1** are analyzed first. During the initial hours, the supplied power demand is notably lower than in the **Base Case**. Due to the interruption of the NGDS supply, the NGFU and dual-fuel unit cannot initially run until gas tankers deliver fuel to them. The stressed fuel condition also affects repair behaviors. In **Case 1**, branch 3-6 is repaired earlier than in the **Base Case**, right after branches 1-2 and 3-5, to swiftly reconnect **Is. 3**, which has been supplied by the NGFU, to the main grid. For the dual-fuel unit, more fuel is transported in two deliveries by gas tankers: one in the 4th TS and another in the 8th TS, each at the tanker's maximum capacity. In addition, GN 6 is restored earlier than in the **Base Case** due to the absence of gas demand of the generating units.

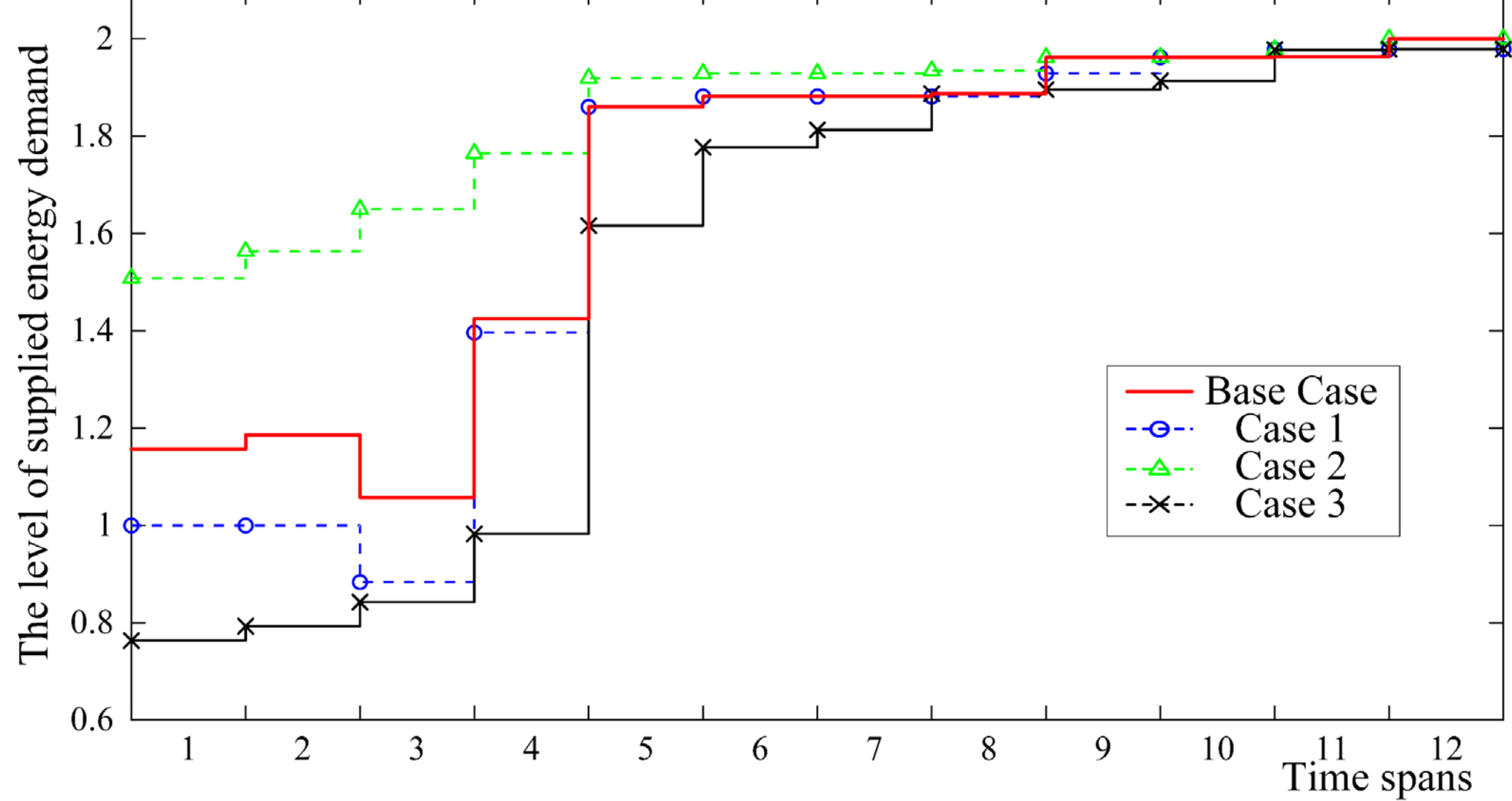

Fig. 9. Comparison of hourly supplied demand across cases.

**Case 2** even outperforms the **Base Case**. With increased initial onsite fuel, generating units can start up immediately and pick up more loads in the initial hours. Throughout the scheduling, the dual-fuel unit alternates between diesel and gas modes. Specifically, it runs in diesel mode in the first two TSs, then switches to gas mode and continues running. A gas tanker refuels the onsite gas storage of the dual-fuel unit in the 8th TS. The unit switches back to diesel mode from the 7th to 9th TSs to meet the peak power demand, which exceeds the capacity of the gas mode but can still be serviced by the diesel mode with a higher capacity. As the onsite diesel is about to be depleted, the dual-fuel unit switches back to gas mode until **Is. 4** reconnects to the main grid at the end of the 11th TS. The RUs exhibit the same behaviors as in the **Base Case**. Additionally, similar to **Case 1**, GN 6 is restored earlier, and less gas demand reduction occurs due to DR compared to the **Bases Case**. The results of **Case 2** highlight the critical importance of preallocating sufficient fuel before disasters and the flexibility of dual-fuel units.

The results of **Case 3** are inferior, yet they particularly underscore the value of gas storage during emergencies. In this case, with further limited gas input, GNs 2, 4, 5, and 6 are restored later than in the **Base Case**, successively picked up after the 3rd TS. During the 5th TS, DFU 36 starts up using fuel from the diesel tanker, while the gas storage intensively supplies gas to the NGDS to meet the gas loads. The gas storage sustains a maximum gas output of 600 $Sm^3/h$ for about four TSs. As gas demand significantly drops, the gas flow from the gas storage decreases to a lower level. Compared to the **Base Case**, the gas storage injects significantly more gas into the NGDS. Additionally, there is less demand reduction due to gas DR. This is primarily because the three DR-related GNs have not yet been restored during the peak time of gas demand. In subsequent periods, as the gas storage continues to support extra supply and the overall load drops, the demand for further DRs also diminishes.

## 5. Conclusion

The growing interdependence between electric power and natural gas systems poses great challenges to their resilience. In this paper, we propose a comprehensive scheduling strategy aimed at enhancing the resilience and restoration capability of the IENDS following disasters. This strategy is designed to synchronize various O&M and logistical processes, addressing critical areas such as fuel supply to generators, power and gas DRs, and repair of damaged facilities, by coordinating the deployment of various emergency resources. Mathematical models are developed to capture the essential characteristics of these processes. Notably, mobile fuel tankers are dispatched to refuel single- and dual-fuel generators, and we construct the model to express the fuel delivery process based on analyzing the fuel exchange relationships among generators, tankers, and other storage facilities. The dispatch of integrated power and gas DRs is modeled using a zonewise approach, which allows for convenient grouping of customers, thereby streamlining the dispatch of DRs to a zonal level. Additionally, the repair process featuring varying efficiencies is incorporated. The developed model enables flexible repair decisions, allowing RUs to either collaborate for enhanced efficiency or work independently across multiple facilities, facilitating a strategic balance between teamwork and individual tasks as needed. Case studies validate the effectiveness of the strategy and also emphasize the value of several measures, including the preallocation of onsite fuel, the flexibility of dual-fuel units, and the utilization of gas storage for emergency supplies.